\documentclass[12pt,notitlepage]{article}
\usepackage{amssymb, latexsym,amsmath,natbib,graphicx,epstopdf,dsfont,outlines,tikz,placeins,amsthm,subfigure,
relsize}
\usepackage[margin=1in]{geometry}

\bibpunct{(}{)}{;}{a}{}{,}

\newcommand{\bPsi}{\boldsymbol\Psi}
\newcommand{\bx}{{\bf X}}
\newcommand{\logit}{\mbox{logit}}
\newcommand{\E}{\mathbb{E}}
\newcommand{\br}{\boldsymbol{r}}

\title{Latent Space Models for Dynamic Networks with Weighted Edges}
\author{Daniel K. Sewell\thanks{Department of Statistics, University of Illinois at Urbana-Champaign, United States}\hspace{0.5pc}\thanks{Corresponding author at: Department of Biostatistics, University of Iowa, 145 N. Riverside Dr., 100 CPHB, Iowa City, IA 52242, United States Tel:(319) 384-1500, E-mail address {\it daniel-sewell@uiowa.edu}} \and Yuguo Chen \footnotemark[1]}
\date{}

\begin{document}



\maketitle
\begin{abstract}
Longitudinal binary relational data can be better understood by implementing a latent space model for dynamic networks.  This approach can be broadly extended to many types of weighted edges by using a link function to model the mean of the dyads, or by employing a similar strategy via data augmentation.  To demonstrate this, we propose models for count dyads and for non-negative real dyads, analyzing simulated data and also both mobile phone data and world export/import data.  The model parameters and latent actors' trajectories, estimated by Markov chain Monte Carlo algorithms, provide insight into the network dynamics.

\vspace{ 2mm}

\noindent
Keywords: Embedding; Markov chain Monte Carlo; Network dynamics; Visualization; Weighted network; Valued dyad
\end{abstract}

\newpage

\section{Introduction}
\label{Introduction}
Representing relational data by networks is extremely useful and widely implemented.   The dyadic relations which compose these networks are viewed as a set of actors and a set of edges between the actors.  The edges can vary in many ways, such as being directed or undirected, static or temporal, binary or weighted.  Binary networks, where between each actor an edge either does or does not exist, are encountered more often in the literature, although many such networks are by nature weighted.  Weighted networks, also referred to as valued networks, consist of actors connected by edges which can take more than two values.  By accounting for the weight, or strength, of the edges, the richness of the data can be better exploited.  
Examples of analyses of real world weighted networks include food webs \citep{krause2003compartments}, gene expression data \citep{zhang2005general}, airline networks \citep{barrat2005effects}, mobile phone networks \citep{onnela2007analysis}, and many more.

Often in binary networks it is of interest to compute various network measures, and recently there has been increasing work in extending these measures to weighted networks.  \cite{opsahl2010node} derived for weighted networks measures for degree, closeness, and betweenness.  \cite{yang2001optimal} derived a method for computing path length in the case of weighted edges.  \cite{opsahl2009clustering} developed a method for analyzing the clustering that exists within a network with weighted edges.  Other interesting works include \cite{kunegis2009slashdot}, which analyzed the case where edges took values in $\{-1,0,1\}$, and \cite{newman2004analysis}, which showed how to model networks whose edges are counts by representing them as multigraphs.  To fully model the network, \cite{krivitsky2012exponential} extended the commonly used exponential random graph model (ERGM) to account for networks whose dyads are counts; \cite{krivitsky2012rank} extended the ERGM to account for networks whose dyads are rankings.

Network data are most often inherently dynamic, even though it is frequently the case that the data are simply aggregated over time into one static network.  Many popular static networks have been extended to longitudinal network data.  Examples of this include the temporal exponential random graph model developed by \cite{hanneke2010discrete} and the separable temporal exponential graph model by \cite{krivitsky2014separable}, the mixed membership stochastic blockmodel for dynamic networks by \cite{xing2010state}, and the latent space model for dynamic networks by several authors including \cite{sarkar2005dynamic}, \cite{SewellChen14}, \cite{morgan2014latent} and \cite{durante2014nonparametric}.

This paper is focused on network data that is dynamic, weighted, and possibly directed.  There are few resources available to the researcher investigating such data.  Most approaches in existence focus on latent space models for dynamic undirected networks.  Latent space models assume the dependence of the network is induced by a set of latent variables.  Such approaches are typically intuitive and have the advantage of producing meaningful visualizations, allowing the researcher to better understand the network structure as well as the behavior of individual actors.

\cite{sarkar2007latent} extended the CODE model of \cite{globerson2004euclidean} for dynamic undirected networks.  This method is an approximate filtering algorithm which models the longitudinal count networks, embedding the actors in a latent space.  This method is not easily generalizable to other sorts of co-occurrence data besides counts, however, and cannot handle directed edges.
\cite{hoff2011hierarchical} described a multilinear model for undirected longitudinal networks.  In this work, Hoff shows how to model undirected edges or ranked edges, where each dyad is an element from a finite ordered set, though it should be feasible to extend their approach to other types of dyads.  \cite{SewellChen14rank} developed a latent space model for directed ranked dynamic networks, where each actor ranks each other actor, although it is not obvious how to extend this approach beyond this specific context.

The remainder of the paper is organized as follows: Section \ref{Models} extends the latent space model for dynamic networks with valued edges.  Section \ref{Estimation} gives a method of estimation.  Section \ref{Scalability} describes an approximation to reduce computational cost for large networks.  Section \ref{Simulations} gives simulation results.  Section \ref{DataAnalysis} gives the results for analyzing Congressional cosponsorship data and world trade data.  Section \ref{Discussion} provides a brief discussion.

\section{Models}
\label{Models}
We assume here that each actor exists within some latent space which can be interpreted as a characteristic space, or a social space.  When actors are closer together in this latent space, the probability of a stronger edge is increased (where a ``stronger edge" means a stronger relationship, though the actual form of this is context specific).

We first introduce some general notation to be used throughout.  Assume we have a set of actors ${\cal N}$ and a set of edges ${\cal E}$.  Let $n=|{\cal N}|$ be the number of actors, and let $Y_t$ be the $n\times n$ adjacency matrix of the observed network at time $t$ whose entries $y_{ijt}$ correspond to the weight of the edge from actor $i$ to actor $j$ for $t\in\{1,2,\ldots,T\}$.  Let ${\bf X}_{it}\in\Re^p$ be the position vector of the $i^{th}$ actor at time $t$ within the $p$ dimensional latent space.  Let ${\cal X}_t$ be the matrix whose $i^{th}$ row is ${\bf X}_{it}$.  Finally, let $\boldsymbol\Psi$ be the vector of unknown parameters (which will vary depending on dyadic type).

As in \cite{sarkar2005dynamic} and \cite{SewellChen14}, we assume the latent actor positions transition according to a Markov process, where the initial distribution is
\begin{equation}
\pi({\cal X}_1|\boldsymbol\Psi)= \prod_{i=1}^n N({\bf X}_{i1}|{\bf 0},\tau^2I_p),
\end{equation}
and the transition equation is
\begin{equation}
\pi({\cal X}_t|{\cal X}_{t-1},\boldsymbol\Psi)=\prod_{i=1}^n N({\bf X}_{it}|{\bf X}_{i(t-1)},\sigma^2I_p) ,
\label{transition}
\end{equation}
for $t=2,3,\ldots,T$, where $I_p$ is the $p\times p$ identity matrix, and $N({\bf x}|\boldsymbol{ \mu},\Sigma)$ denotes the multivariate normal probability density function with mean $\boldsymbol\mu$ and covariance matrix $\Sigma$ evaluated at ${\bf x}$.  While this is the latent dependence structure used throughout the remainder of the paper, other dependence structures could be defined, such as the latent path model given by \cite{morgan2014latent}.

\begin{figure}[t]
\centering
\vspace{-3pc}
\includegraphics[scale=0.6]{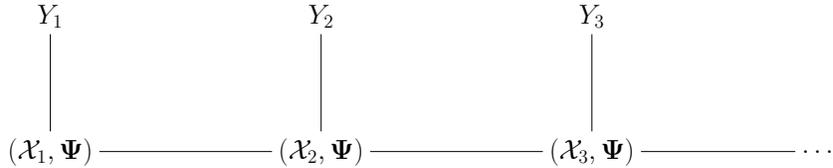}\vspace{-21pc}
\caption{Illustration of the dependence structure for the latent space model.  $Y_t$ is the observed graph, ${\cal X}_t$ is the unobserved latent actor positions, and $\boldsymbol\Psi$ is the vector of model parameters.
}
\label{dep_struct}
\end{figure}

In most dynamic network models it is assumed that the dependence structure of the network is fully induced by the latent positions of the actors.   This assumption, along with the Markovian properties of the latent positions, leads to the state space temporal dependence structure given in Figure \ref{dep_struct}, as well as the conditional independence of each dyad within a time period.  The ranked networks of the form analyzed by \cite{krivitsky2012rank} and \cite{SewellChen14rank} are a counter example of where there is an extra dependency constraint in the data, but we will not discuss further these rare data types.  What remains then is to derive an appropriate conditional likelihood function, $\pi(Y_1,\ldots,Y_T|{\cal X}_1,\ldots,{\cal X}_T,\boldsymbol\Psi)=\prod_{t=1}^T\prod_{i\neq j}\pi(y_{ijt}|{\cal X}_t,\boldsymbol\Psi)$.  

Most latent space approaches have the conditional likelihood constructed by writing the logit of the edge probability as a linear form of covariates and a function of the latent variables, i.e., $\logit(\pi(y_{ijt}|\cdot))=\boldsymbol{\alpha}'{\textbf w}_{ijt} + f_{\bPsi}(\bx_{it},\bx_{jt})$, where $\boldsymbol{\alpha}$ is a vector of unknown parameters, ${\textbf w}_{ijt}$ is a vector of dyad specific covariates, and $f_{\bPsi}:\Re^p\times\Re^p\rightarrow \Re$ is a function taking as its arguments two actors' latent variables.  Our generalization of this has the basic form
\begin{equation}
g(\E(y_{ijt}))=\boldsymbol{\alpha}'{\textbf w}_{ijt} + f_{\bPsi}(\bx_{it},\bx_{jt}),
\label{link1}
\end{equation}
for some link function $g$.  We can utilize the same types of link functions found in generalized linear mixed models.  For example if our dyads are in the form of continuous data, we may set $g$ to be the identity; this may arise in, for instance, proximity networks \citep[see, e.g.,][]{olguin2009capturing}, where the distance between individuals is recorded on a regular basis.  The common case of modeling binary dyads through the logit link function is yet another example.  In Section \ref{Counts} we will go into detail for the context of count data, using a log link function.

In some cases, however, the dyads cannot be modeled directly through a link function as in (\ref{link1}).  Instead we can introduce additional latent variables, and then adopt a similar strategy.  For example, we may consider a zero inflated model.  
The zero inflated model is a two component mixture model, where one could introduce additional latent indicator variables which determine whether the observation is coming from the component which is a point mass at zero or the component that has some other density function $\pi^*$ (e.g., $\pi^*$ is the Poisson density).  We could then model $g(\E_{\pi^*}(y_{ijt}))$ as in (\ref{link1}).  This situation may arise in large sparse weighted network data, such as company wide email count networks.  Zero-inflated models are certainly not the only possibility of this type of data augmentation, as we will see in Section \ref{NonNegativeContinuousEdges}.

For the remainder of the paper we will focus on count data and non-negative continuous edges.  We will furthermore utilize the conditional likelihood given by \cite{SewellChen14}, determined by
\begin{equation}
f_{\bPsi}(\bx_{it},\bx_{jt})=\beta_{IN}\left(1-\frac{d_{ijt}}{r_j}\right)+\beta_{OUT}\left( 1- \frac{d_{ijt}}{r_i} \right),
\label{link2}
\end{equation}
where $d_{ijt}=\|{\bf X}_{it}-{\bf X}_{jt}\|$ is the distance between actors $i$ and $j$ at time $t$ within the latent space, and $\boldsymbol{r}=(r_1,r_2,\ldots,r_n)$ is a vector of positive actor specific parameters constrained such that $\sum_{i=1}^nr_i=1$ for model identifiability.  

Each $r_i$ can be thought of as the $i^{th}$ actor's social reach.  That is, a larger value of $r_i$ implies that it is more likely for an edge, either $y_{i\cdot t}$ or $y_{\cdot it}$, to take a larger value.  These $r_i$'s also hold a geometric interpretation within the latent space, specifically a radius.  For example, in the context of binary networks, this radius can be understood to imply that actors inside of each others' radii have a greater than $1/2$ probability of an edge, and actors are outside of each other's radii have a smaller than $1/2$ probability of an edge.  The coefficients $\beta_{IN}$ and $\beta_{OUT}$ can help in understanding the global structure of the network, insofar as telling us whether activity (tendency to send stronger edges) or popularity (tendency to receive stronger edges) is more important in forming high strength edges.  Specifically, $\beta_{IN}>\beta_{OUT}$ implies popularity is more important than activity in the edge formation process, and $\beta_{OUT}>\beta_{IN}$ implies the opposite.  If the edges are undirected, then setting $\mathbb{P}(y_{ijt}|\cdot)=\mathbb{P}(y_{jit}|\cdot)$ is equivalent to constraining $\beta_{IN}=\beta_{OUT}$.  See \cite{SewellChen14} for more details on parameter interpretation.

\subsection{Counts}
\label{Counts}
A commonly encountered dyadic type which can be modeled by (\ref{link1}) is where $y_{ijt}$ is a count.  This context may exist in the form of counting the number of phone calls, the number of emails, the number of cosponsored legislative bills, the number of passengers or of flights in airline data, etc.  We can use the canonical link for a Poisson random variable to determine the likelihood function in the following way:
\begin{equation}
\mathbb{P}(y_{ijt}|{\cal X}_t,\boldsymbol\Psi)= \frac{\lambda_{ijt}^{y_{ijt}}\exp(-\lambda_{ijt})}{y_{ijt}!}, \hspace{2pc} y_{ijt}=0,1,2,\ldots
\label{count_Lik1}
\end{equation}
where
\begin{equation}
\log(\lambda_{ijt})=\beta_{IN}\left(1-\frac{d_{ijt}}{r_j}\right)+\beta_{OUT}\left(1-\frac{d_{ijt}}{r_i}\right).
\label{count_Lik2}
\end{equation}
Here $\boldsymbol\Psi=(\beta_{IN},\beta_{OUT},\boldsymbol{r},\tau^2,\sigma^2)$ is the vector of parameters.  Thus the likelihood is
\begin{equation}
\mathbb{P}(Y_1,Y_2,\ldots,Y_T|{\cal X}_1,{\cal X}_2,\ldots,{\cal X}_T,\boldsymbol\Psi)= \prod_{t=1}^T\prod_{i\neq j} \frac{\lambda_{ijt}^{y_{ijt}}\exp(-\lambda_{ijt})}{y_{ijt}!}.
\end{equation}

\subsection{Non-Negative Continuous Edges}
\label{NonNegativeContinuousEdges}
Here we consider the case of non-negative continuous real valued edges.  These types of networks can occur in many biological contexts, in economic contexts, in length of phone calls, etc.  The latent space framework provides a natural way to think about such a weighted network, in that we can consider two actors with large weighted edges between them as very close in the latent space, and two actors with smaller weighted edges as more separated within the latent space. 

By embedding the network into a latent space we can better differentiate between zero valued edges. Consider as an example a longitudinal sequence of social networks, where the dyadic variable measured is the amount of time two individuals spent speaking with each other: Suppose at a particular time, person $i$ has a weighted edge of zero with two others, persons $j$ and $k$. Now persons $i$ and $j$ are potential friends though they have not currently met; meanwhile, persons $i$ and $k$ know each other already and strongly dislike each other. In both cases the measured edges between $i$ and $j$ and between $i$ and $k$ will be the same (zero), but we can differentiate them in two ways.  First and foremost we can compare edge probabilities (e.g., $\mathbb{P}(y_{ij}=0)<<\mathbb{P}(y_{ik}=0)$).  Second, viewing the latent variables as unobserved actor attributes, we can determine the dissimilarity between each pair (e.g., $d_{ij}<<d_{ik}$).  The key point here is that we are using all the data, not just the data from the pairs $(i,j)$ and $(i,k)$, to learn more about such observed zeros; i.e., we are letting all dyads help inform us as to the position of each actor within the latent space.  This can be better understood by considering if $i$ and $j$ have many links to the same actors, then the geometric constraints within the latent space imply that $i$ and $j$ will be close together, whereas the same would not be true if, say, $i$ and $k$ do not have many links to the same actors.

Network data with non-negative continuous edges is a context where there is not an obvious link function $g$ to be applied to the mean of $y_{ijt}$, but by introducing an additional latent variable we can adopt a similar strategy.  In particular, we apply a tobit model when formulating the likelihood function, 
letting $y_{ijt}= y_{ijt}^*1_{\{ y_{ijt}^*>0 \}}$, where $1_{\{\cdot\}}$ is the indicator function and $y_{ijt}^*$ is a continuous normal random variable.  This type of approach may be most appropriate when the weighted dyads we observe are really proxies for some underlying relationship between the two actors, but we can only observe the effects from positive relationships (e.g., length of phone calls can only serve as a proxy for a relationship between friends, and not between enemies).  We then apply (\ref{link1}) to the latent variables $y_{ijt}^*$, letting $g$ be the identity function, obtaining
\begin{equation}
y_{ijt}^*=\beta_{IN}\left(1-\frac{d_{ijt}}{r_j}\right)+\beta_{OUT}\left(1-\frac{d_{ijt}}{r_i}\right) + \epsilon_{ijt},
\label{ystar}
\end{equation}
\begin{equation}
\epsilon_{ijt}\mathlarger{|} ({\cal X}_t,\boldsymbol\Psi) \stackrel{iid}{\sim} N(0,\gamma^2).
\end{equation}
With this we have
\begin{equation}
\pi(y_{ijt}|{\cal X}_t, \boldsymbol\Psi)= \left[ N(y_{ijt}|\mathbb{E}(y_{ijt}^*|{\cal X}_t,\boldsymbol\Psi),\gamma^2)\right]^{1_{\{y_{ijt}>0\}}}\left[ 1-\Phi\left( \frac{\mathbb{E}(y_{ijt}^*|{\cal X}_t,\boldsymbol\Psi)}{\gamma} \right) \right]^{1_{\{y_{ijt}=0\}}},
\label{tobit_Lik}
\end{equation}
where $\Phi$ is the standard normal cumulative distribution function, and $\mathbb{E}(y_{ijt}^*|{\cal X}_t,\boldsymbol\Psi)=$\linebreak$\beta_{IN}\left(1-d_{ijt}/r_j\right)+\beta_{OUT}\left(1-d_{ijt}/r_i\right)$ is the conditional expectation of $y_{ijt}^*$.  The vector of parameters is now supplemented by $\gamma^2$ such that $\boldsymbol\Psi=(\beta_{IN},\beta_{OUT},\gamma^2,\boldsymbol{r}, \tau^2,\sigma^2)$.  Since the $\epsilon_{ijt}$'s are conditionally i.i.d., we have that the observation equation is
\begin{equation}
\mathbb{P}(Y_1,Y_2,\ldots,Y_T|{\cal X}_1,{\cal X}_2,\ldots,{\cal X}_T,\boldsymbol\Psi)=\prod_{t=1}^T\prod_{i\neq j} \pi(y_{ijt}|{\cal X}_t,\boldsymbol\Psi).
\end{equation}

\section{Estimation}
\label{Estimation}
To obtain estimates of the latent space positions and of the unknown parameters, we sample via a Markov chain Monte Carlo (MCMC) algorithm from the posterior
\begin{equation}
\pi({\cal X}_1,{\cal X}_2,\ldots,{\cal X}_T,\boldsymbol\Psi|Y_1,Y_2,\ldots,Y_T).
\end{equation}
The general strategy is to find reasonable estimates of the latent positions and of the model parameters to initialize the chain, and then use a Metropolis-Hastings (MH) within Gibbs sampling to obtain the posterior samples.

The prior for $\boldsymbol{r}$ was a Dirichlet distribution.  The priors for $\tau^2$, $\sigma^2$ and, in the case of continuous data, $\gamma^2$ were chosen to be inverse gamma (IG), as these distributions are conjugate for $\tau^2$ and $\sigma^2$. The shape and scale parameters for $\tau^2$ were set to be equal to $2+\delta$ and $(1+\delta)\tau_0^2$ respectively for some small $\delta$ (we used 0.05) and some positive constant $\tau_0^2$, and were similarly set for $\sigma^2$ and $\gamma^2$.  With this parameterization, the prior variances of $\tau^2$, $\sigma^2$ and $\gamma^2$ are kept large.  The prior set on $\beta_{IN}$ was a normal distribution with mean $\nu_{IN}$ and (large) variance $\xi_{IN}$, and similarly for $\beta_{OUT}$.  

In some cases, by taking a preliminary look at the data we can set these hyperparameters to reasonable values; specifically, we may match the prior means to the initialized values given in the next section.  This ``first glance" allows us to form our prior beliefs about the scale of the parameters, as well as about the individual actor effects.  This idea follows the same underlying concept as empirical Bayes methodology, in that we are (albeit to a small degree in comparison to standard empirical Bayes methods) using the data to construct the hyperparameters.  This idea was used in \cite{SewellChen14} to good effect, and also yielded good results in our analyses.  For some other parameters there is not an obvious way in which we can set the hyperparameters in this fashion.  As will be described in the simulation study, however, the results were not sensitive to the selection of these hyperparameter values.

\subsection{Initialization}
\label{Initialization}
We initialized the radii as
\begin{equation}
r_i=\frac{\sum_{t=1}^T \sum_{j\neq i}(y_{ijt}+y_{jit})/2 }
{\sum_{t'=1}^T \sum_{i'=1}^n\sum_{j'\neq i'} y_{i'j't'}}.
\label{init_weights}
\end{equation}
The Dirichlet hyperparameters for $\br$ were set to be equal to these initial estimates.  Doing so sets the prior expected value of $r_j$ to be the initial estimate of $r_j$, which reflects our prior intuition; additionally, as each $\alpha_j$ would be small (averaging $1/n$), the prior variance for each $r_i$ will be large (leading to a ``flat" prior).  

To find initial latent positions, we implemented the generalized multidimensional scaling algorithm (GMDS), as described in \cite{sarkar2005dynamic}.  GMDS starts by taking a distance matrix at time 1 and performing classical multidimensional scaling.  Then, for each subsequent time period $t$, $t=2,3,\ldots,T$, GMDS balances the position matrix from the previous time point with the classical multidimensional scaling result obtained from the distance matrix at time $t$.

The original distance matrices can be found in a number of ways, but we offer the following suggestion.  We treated the data as binary, where
\begin{equation}
y^{(binary)}_{ijt}=\left\{\begin{array}{cc} 1 & \mbox{ if } y_{ijt} > 0 \\ 0 & \mbox{ otherwise.} \end{array} \right.
\end{equation}
We then computed each distance $d_{ijt}$ according to
\begin{equation}
d_{ijt}=\left\{
\begin{array}{ll}
\frac{1}{2}\min\{ r_i,r_j \} & \mbox{ if } y^{(binary)}_{ijt}=y^{(binary)}_{jit}=1 \\
\frac12(r_i+r_j) & \mbox{ if } y^{(binary)}_{ijt}+y^{(binary)}_{jit}=1 \\
\frac32(r_i+r_j)& \mbox{ if } y^{(binary)}_{ijt}=y^{(binary)}_{jit}=0.
\end{array} \right.
\end{equation}
The general idea here is that positive edges indicate a closeness between the actors; using the radii as measures of closeness accounts for the individual effects as well as keeps the distances on the same scale as the radii, as would seem reasonable based on (\ref{link2}).  With the $T$ distance matrices computed, we can then implement GMDS to obtain initial latent positions.

The initial estimate for $\tau^2$ was computed (using the initial estimates of ${\cal X}_1$) as
\begin{equation}
\frac{1}{np}\sum_{i=1}^n\|{\bf X}_{i1}\|^2.
\label{tau2}
\end{equation}
In our analyses we also used this value to determine the hyperparameter $\tau_0^2$, the prior mean of $\tau^2$.  

We found that the initial value of $\sigma^2$ did not make a noticeable difference in the performance, nor did the value of $\sigma^2_0$.  Similarly, the initial estimates for $\gamma^2$, $\beta_{IN}$ and $\beta_{OUT}$ and the values of their hyperparameters did not significantly affect the number of iterations required to reach convergence.

\subsection{Posterior Sampling}
\label{PosteriorSampling}
We implement a MH within Gibbs sampling scheme.  The algorithm is
\begin{description}
\item[0.] Set the initial values of the latent positions and parameters as given in Section 3.1.\vspace{-1pc}
\item[1.] For $t=1,2,\ldots,T$ and for $i=1,2,\ldots,n$, draw ${\bf X}_{it}$ via MH. \vspace{-1pc}
\item[2.] Draw $\tau^2$ from $\pi(\tau^2|{\cal X}_1)$.\vspace{-1pc}
\item[3.] Draw $\sigma^2$ from $\pi(\sigma^2|{\cal X}_1,{\cal X}_2,\ldots,{\cal X}_T)$.\vspace{-1pc}
\item[4.] Draw $\boldsymbol{r}$ via MH.\vspace{-1pc}
\item[5.] Draw $\beta_{IN}$ via MH.\vspace{-1pc}
\item[6.] Draw $\beta_{OUT}$ via MH.\vspace{-1pc}
\item[] \hspace{-2pc} {\it If data is non-negative continuous}\vspace{-1pc}
\item[7.] Draw $\gamma^2$ via MH.\vspace{-1pc}
\item[]Repeat steps 1-7.\vspace{0pc}
\end{description}

The full conditional distributions needed for steps 2-7 are given in the Appendix.
Regarding the proposal distributions, ${\bf X}_{it}$, $\beta_{IN}$, and $\beta_{OUT}$ can come from a symmetric proposal (e.g., normal random walk).  For $\gamma^2$, however, some asymmetric proposal such as a log-normal (what we used in our analyses) or an inverse gamma distribution ought to be used to ensure positive valued proposals; this asymmetric proposal will then need to be accounted for in the acceptance probability.  Because of the constraint on $\boldsymbol{r}$, a Dirichlet proposal is suggested for the radii, which also will be an asymmetric proposal.  Suggested parameters for this Dirichlet proposal are $\kappa\boldsymbol{r}^{curr}$, where $\boldsymbol{r}^{curr}$ are the current values for $\boldsymbol{r}$ and $\kappa$ is some large value.

One final note is that, as is the case for any such latent space model, the posterior is invariant under rotations, reflections and translations of the latent positions ${\cal X}_1,{\cal X}_2, \ldots,{\cal X}_T$.  In order to make the MCMC iterations comparable, after each iteration of steps 1-7 we perform a Procrustes transformation on the $nT\times p$ matrix $({\cal X}_1',\ldots,{\cal X}_T')'$.  The Procrustean transformation finds a set of rotations, reflections and translations to minimize the difference between a given matrix and some target matrix.  In our analyses, we constructed the target matrix from the initialized latent position trajectories.

\section{Scalability}
\label{Scalability}
The MCMC algorithm of Section \ref{PosteriorSampling} can handle many data sets, including the two that are fully described in Section \ref{DataAnalysis}.  However, in cases where the network is very large, the MCMC algorithm may prove to be too slow to be viable.  For static binary latent space network models, \cite{raftery2012fast} described a method for approximating the log likelihood using case-control principles.  \cite{SewellChen14} also used this method for binary dynamic latent space network models, adapting it slightly to allow for missing data.  Here we extend this for models whose dyads can be described by an exponential family of distributions.

For the MCMC algorithm, the MH steps required in updating the latent positions, $\boldsymbol{r}$, $\beta_{IN}$, $\beta_{OUT}$ and other likelihood related parameters (e.g., $\gamma^2$ in the case of non-negative real dyads) all require $O(Tn^2)$ terms to be summed.  In this discussion we will assume here that a non-relationship between two actors implies that $y_{ijt}=0$, otherwise $y_{ijt}$ is some positive value; the principles discussed next ought to hold even if this is not the case.  Generalizing the approximation method first proposed by \cite{raftery2012fast}, we can reduce this computational cost to $O(Tn)$.

Suppose that, conditional on the latent positions, the $y_{ijt}$'s are independent with
\begin{equation}
\pi(y_{ijt}|\cdot)=h(y_{ijt})\exp(\boldsymbol\eta_{ijt}'{\bf T}(y_{ijt})+A(\boldsymbol\eta_{ijt})),
\label{expFam}
\end{equation}
where $\boldsymbol\eta_{ijt}$ is a vector valued function of $({\cal X}_t,\boldsymbol\Psi)$ and ${\bf T}(y_{ijt})$ is a vector of sufficient statistics.  Then we can rewrite the loglikelihood of $(Y_1,\ldots,Y_T)$ as
\begin{align}\nonumber
\ell({\cal X}_1,\ldots,{\cal X}_T,\boldsymbol\Psi)= & \sum_{t=1}^T \sum_{i=1}^n\left[
\sum_{j:y_{ijt}>0}\big(\boldsymbol\eta_{ijt}'{\bf T}(y_{ijt}) + A(\boldsymbol\eta_{ijt})\big) \right. &\\
& \left. +
\sum_{j:y_{ijt}=0}\big(\boldsymbol\eta_{ijt}'{\bf T}(y_{ijt}) + A(\boldsymbol\eta_{ijt})\big)\right]
+\mbox{constant}.
\label{genLogLik1}
\end{align}
It is reasonable to assume that as the network gets larger and larger, the number of edges of each node does not grow at the same rate (i.e., the network gets sparser).  Hence we make the assumption that either the maximum degree is fixed or is of $o(n)$.  If this is the case, then we can, for each $i$ and $t$, take a subsample $\{j_k\}_{k=1}^{N_{i,t,0}}$ from the set $\{j:y_{ijt}=0\}$ and use a simple Monte Carlo estimate of the final summation of (\ref{genLogLik1}) to reduce the computational cost to linear with respect to $n$.  Then the approximation we use of the log likelihood is
\begin{align}\nonumber
\ell({\cal X}_1,\ldots,{\cal X}_T,\boldsymbol\Psi)\approx &
\sum_{t=1}^T \sum_{i=1}^n\left[
\sum_{j:y_{ijt}>0}\big(\boldsymbol\eta_{ijt}'{\bf T}(y_{ijt}) + A(\boldsymbol\eta_{ijt})\big) \right.&\\
&\left.+
\frac{n_{i,t,0}}{N_{i,t,0}}
\sum_{k=1}^{N_{i,t,0}}\big(\boldsymbol\eta_{ij_kt}'{\bf T}(y_{ij_kt}) + A(\boldsymbol\eta_{ij_kt})\big)\right]
+\mbox{constant},
\label{genLogLik2}
\end{align}
where $n_{i,t,0}=|\{j:y_{ijt}=0\}|$.  In most cases, ${\bf T}(y_{ijt})=0$ if $y_{ijt}=0$ and hence the above can be simplified such that the second summation is only $\frac{n_{i,t,0}}{N_{i,t,0}}
\sum_{k=1}^{N_{i,t,0}}A(\boldsymbol\eta_{ij_kt})$.  Also, there could potentially be multiple methods of selecting the subsequences $\{j_k\}_{k=1}^{N_{i,t,0}}$; see \cite{raftery2012fast} for more details.

For $T=1$ and $y_{ijt}\in\{0,1\}$, this leads to Raftery et al.'s approximation.  For the context presented in Section \ref{Counts}, we can approximate the log likelihood as
\begin{equation}
\ell({\cal X}_1,\ldots,{\cal X}_T,\boldsymbol\Psi)\approx
\sum_{t=1}^T \sum_{i=1}^n\left[
\sum_{j:y_{ijt}>0}\Big(
y_{ijt}\log(\lambda_{ijt}) +\lambda_{ijt}
\Big)+
\frac{n_{i,t,0}}{N_{i,t,0}}
\sum_{k=1}^{N_{i,t,0}}\lambda_{ij_kt}
\right] +\mbox{constant}.
\end{equation}
For the context presented in Section \ref{NonNegativeContinuousEdges}, we can approximate the log likelihood as
\begin{align}\nonumber
\ell({\cal X}_1,\ldots,{\cal X}_T,\boldsymbol\Psi)\approx &
\sum_{t=1}^T \sum_{i=1}^n\left\{
\sum_{j:y_{ijt}>0}\left[
-\frac{1}{2}\log(\gamma^2)-\frac{1}{2\gamma^2}(y_{ijt}-\mathbb{E}(y_{ijt}^*|{\cal X}_t,\boldsymbol\Psi))^2
\right]\right.& \\
\left. \right.&\left.+\frac{n_{i,t,0}}{N_{i,t,0}}
\sum_{k=1}^{N_{i,t,0}}\log\Big( 1-\Phi( \mathbb{E}(y_{ijt}^*|{\cal X}_t,\boldsymbol\Psi)/\gamma ) \Big)
\right\} + \mbox{constant}.&
\end{align}.

An interesting point is that if we assume that the network becomes more sparse as $n$ grows, then it may be more appropriate to utilize a zero-inflated model, such as was mentioned in Section \ref{Models}.  Suppose we can augment the data by component indicator variables $z_{ijt}\in\{1,2\}$, such that $\pi(y_{ijt}|z_{ijt}=1,\cdot)=\delta(y_{ijt})$, $\pi(y_{ijt}|z_{ijt}=2,\cdot)$ can be constructed according to (\ref{expFam}), where $\delta$ is the Dirac delta function.  Letting $\pi(z_{ijt}=1)=\alpha$, we can then write the approximated complete log likelihood (i.e., $\pi(Y_1,\ldots,Y_T,Z_1,\ldots,Z_T|\cdot)$) as
\begin{align}\nonumber
&\ell({\cal X}_1,\ldots,{\cal X}_T,\boldsymbol\Psi)&\\ \nonumber
=& \sum_{t=1}^T\sum_{i=1}^n 
\left\{
\sum_{j:y_{ijt}>0}
\left[
\log(1-\alpha)+\boldsymbol\eta_{ijt}'{\bf T}(y_{ijt}) + A(\boldsymbol\eta_{ijt})
\right] \right.&\\
&\left.+
\frac{n_{i,t,0}}{N_{i,t,0}}
\sum_{k=1}^{N_{i,t,0}}
\left[
1_{\{z_{ij_kt}=1 \}}\log(\alpha) + 1_{\{z_{ij_kt}=2 \}}\Big(\log(1-\alpha)+ \boldsymbol\eta_{ij_kt}'{\bf T}(y_{ij_kt}) + A(\boldsymbol\eta_{ij_kt})\Big)
\right]
\right\} + \mbox{constant},
\end{align}
where $[Z_t]_{ij}=z_{ijt}$.  Since the $z_{ijt}$'s are nuisance parameters, we need not sample all of them in the Gibbs sampler, but rather only the $z_{ijt}$'s corresponding to each of the $n$ subsequences $\{j_k\}_{k=1}^{N_{i,t,0}}$, thus maintaining the computational cost of $O(Tn)$.

\section{Simulations}
\label{Simulations}
We analyzed simulated data for both count data and non-negative continuous data.  In each case we simulated twenty data sets where the number of actors was 100 and the number of time points was 10.  The values used in these simulations, given in the next two sections, were chosen to create data that was similar to the real data we analyzed.

\subsection{Simulated Count Data}
\label{SimulatedCountData}
For each of the twenty simulations, the parameter values were set at $\beta_{IN}=3$ and $\beta_{OUT}=1$.  The transition variance was set to be $\sigma^2=1\times10^{-6}$.  The latent positions at time 1 were drawn from a mixture of 10 normals with equal mixture component weights, where the cluster means were drawn randomly from a multivariate normal distribution with mean zero and covariance $((9/10)\cdot2\times 10^{-5}) I_p$, and where the cluster covariances were $((1/10)\cdot2\times 10^{-5}) I_p$, and $p=2$ is the dimension of the latent space.  After the initial latent positions ${\cal X}_1$ were drawn, the radii $\boldsymbol{r}$ were drawn from a Dirichlet distribution whose $i^{th}$ parameter was equal to $n(1/\|{\bf X}_{i1}\|)/\max_k\{ 1/\|{\bf X}_{k1}\| \}$, thus giving those centrally located actors a large individual effect, which reflects a reality.  Subsequent latent positions ${\cal X}_t$, $t\geq2$, were drawn according to (\ref{transition}).  The adjacency matrices $Y_1$ to $Y_T$ were then generated according to (\ref{count_Lik1}) and (\ref{count_Lik2}).  The mean proportion of edges that were positive over the simulations was 0.56, ranging from 0.34 to 0.69.

For each simulation we drew $\sigma_0^2$ from $U(1\times10^{-4},1\times10^{-2})$, where $U(a,b)$ is the uniform distribution over the interval $(a,b)$.  Both hyperparameters $\nu_{IN}$ and $\nu_{OUT}$ were for each simulation drawn from $U(1,15)$; $\xi_{IN}$ and $\xi_{OUT}$ were set to be 1,000.

To evaluate the simulation results, we compared the estimates of the coefficients $\beta_{IN}$ and $\beta_{OUT}$ with the truth, evaluated the pseudo $R^2$, and evaluated the pairwise ratios of estimated distances to true distances corresponding to the latent positions.  The pseudo $R^2$
value is the deviance based pseudo $R^2$ for count data found, and recommended, in \cite{cameron1996r}.  This is calculated as
\begin{equation}
R^2=\frac{\sum_{t=1}^T\sum_{i\neq j}y_{ijt}\log(\hat{\lambda}_{ijt}/\bar{y})-(\hat{\lambda}_{ijt}-\bar{y})}{\sum_{t'=1}^T\sum_{i'\neq j'}y_{i'j't'}\log(y_{i'j't'}/\bar{y})},
\end{equation}
where $\bar{y}=\sum_{t=1}^T\sum_{i\neq j}y_{ijt}$ and $\hat{\lambda}_{ijt}$ is found by plugging in the posterior mean estimates in (\ref{count_Lik2}).
To clarify what is meant by the distance ratios, note that for each simulation there are $Tn(n-1)/2$ distances within the latent space.  We calculate all these pairwise distances using the posterior mean latent positions as well as using the true latent positions.  So for each simulation we can plot a curve corresponding to the distribution of these ratios.  We would hope for this curve to be narrow and centered at 1.

The posterior mean estimate, averaged over the ten simulations, for $\beta_{IN}$ ($\beta_{OUT}$) whose true value was 3 (1), was 2.95 (1.01), ranging from 2.84 to 3.01 (0.969 to 1.06).  
The pseudo $R^2$ values' average was 0.930, ranging from 0.908 to 0.944, implying that the posterior means fit the data well.  
The distributions of the ratios of pairwise distances are given in Figure \ref{countDistDists}, where each curve corresponds to a simulation.  From this figure we see that the picture we obtain from the estimated latent space is close to the true latent space, up to a rotation/translation, for all but perhaps one simulation; this outlying simulation still yields a narrow distribution, implying that the picture of the latent space is close to the truth up to a scaling factor.  In each simulation we are satisfied with the results; considering that $\sigma^2_0$, $\nu_{IN}$, and $\nu_{OUT}$ were randomized in each case, we can conclude that the results are not sensitive to these hyperparameters.

\subsection{Simulated Continuous Data}
\label{SimulatedContinuousData}
For each of the twenty simulations, the parameter values were set at $\beta_{IN}=3$, $\beta_{OUT}=1$, $\gamma^2=4$, and $\sigma^2=1\times10^{-6}$.  The latent positions at time 1 were drawn from a mixture of 10 normals with equal mixture component weights, where the cluster means were drawn randomly from a multivariate normal distribution with mean zero and covariance $((9/10)\cdot2\times 10^{-5}) I_p$, and where the cluster covariances were $((1/10)\cdot2\times 10^{-5}) I_p$, and $p=2$ is the dimension of the latent space.  After the initial latent positions ${\cal X}_1$ were drawn, the radii $\boldsymbol{r}$ were drawn from a Dirichlet distribution whose $i^{th}$ parameter was equal to $n(1/\|{\bf X}_{i1}\|)/\max_k\{ 1/\|{\bf X}_{k1}\| \}$.  Subsequent latent positions ${\cal X}_t$, $t\geq2$, were drawn according to (\ref{transition}).  The adjacency matrices $Y_1$ to $Y_T$ were then constructed by generating $y_{ijt}^*$ according to (\ref{ystar}) and letting $y_{ijt}=y_{ijt}^*1_{\{y_{ijt}^*>0\}}$.  The mean proportion of edges that were positive over the simulations was 0.480, ranging from 0.293 to 0.590.

For each simulation we drew $\sigma^2_0$ from $U(1\times 10^{-4},1\times10^{-2})$, $\gamma^2_0$ from $U(1,5)$, and both $\nu_{IN}$ and $\nu_{OUT}$ from $U(1,15)$; both $\xi_{IN}$ and $\xi_{OUT}$ were set to be 1,000.

To evaluate the simulation results, we compared the estimates of the coefficients $\beta_{IN}$ and $\beta_{OUT}$ with the truth, evaluated the pseudo $R^2$, and evaluated the pairwise ratios of estimated distance to true distance.  In this context of continuous non-negative data, we used the pseudo $R^2$ value recommended in \cite{veall1994practitioners}, originally derived by \cite{mckelvey1975statistical}.  This is calculated as
\begin{equation}
R^2=\frac{\sum_{t=1}^T\sum_{i\neq j}(\hat{y}_{ijt}^*-\hat{\bar{y}}^*)^2}{\sum_{t'=1}^T\sum_{i'\neq j'}(\hat{y}_{i'j't'}^*-\hat{\bar{y}}^*)^2 + Tn(n-1)\widehat{\gamma}^2},
\end{equation}
where $\hat{y}_{ijt}^*=\hat{\beta}_{IN}(1-\hat{d}_{ijt}/\hat{r}_j)+\hat{\beta}_{OUT}(1-\hat{d}_{ijt}/\hat{r}_i)$ and $\hat{\bar{y}}^*=1/(Tn(n-1))\sum_{t=1}^T\sum_{i\neq j}\hat{y}_{ijt}^*$.  The $\hat{}$ symbol over the model parameters implies the posterior mean estimate.

\begin{figure}
\centering
\subfigure[]
{
\includegraphics[width=5cm,height=5cm]{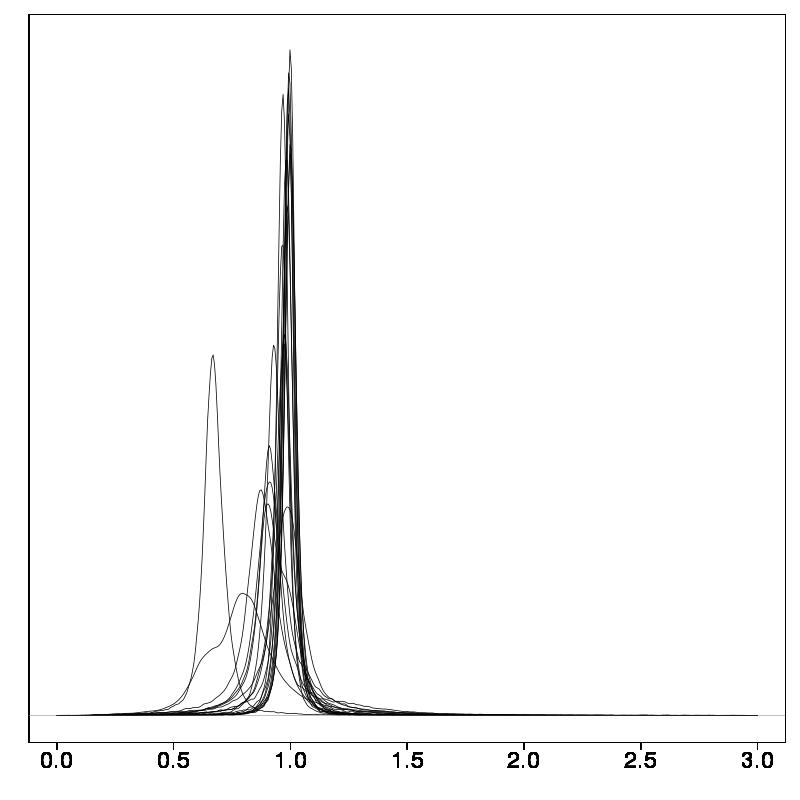}
\label{countDistDists}
}
\subfigure[]
{
\includegraphics[width=5cm,height=5cm]{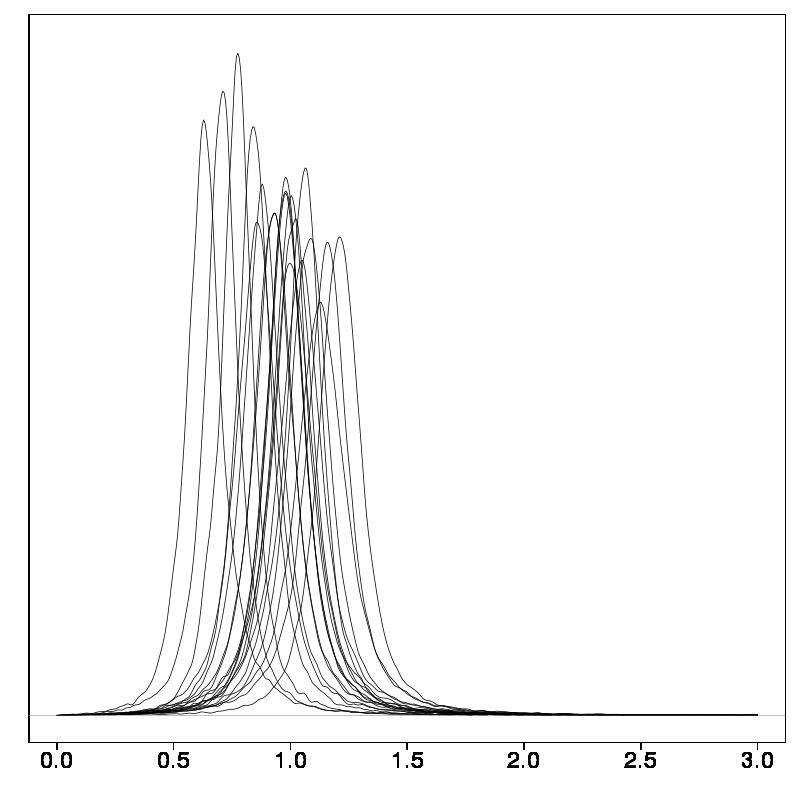}
\label{tobitDistDists}
}
\caption{Left and right columns correspond to count data and non-negative continuous data respectively: (a)-(b) boxplot of posterior means of $\beta_{IN}$ and $\beta_{OUT}$; (c)-(d) boxplot of pseudo-$R^2$; (e)-(f) distributions of pairwise ratios of estimated distances and true distances}
\end{figure}

The posterior mean estimate, averaged over the twenty simulations, for $\beta_{IN}$ ($\beta_{OUT}$) whose true value was 3 (1), was 2.96 (1.01), ranging from 2.63 to 3.12 (0.938 to 1.11).  The pseudo $R^2$ values' average was 0.854, ranging from 0.660 to 0.982, implying that the posterior means fit the data well.  The distributions of the ratios of pairwise distances are given in Figure \ref{tobitDistDists}, where each curve corresponds to a simulation.  Nearly all of these are narrow and centered near one, and all seem narrow, implying that the picture we obtain from the estimated latent space is close to the true latent space up to a rotation/translation and sometimes a small scalar.  In each simulation we are satisfied with the results; considering that $\sigma^2_0$, $\gamma^2_0$, $\nu_{IN}$, and $\nu_{OUT}$ were randomized in each case, we can conclude that the results are not sensitive to these hyperparameters.

\section{Data Analysis}
\label{DataAnalysis}
\subsection{Friends and Family Data}
\label{FFData}
We consider the Friends and Family data collected by the MIT Human Dynamics Lab \citep{aharony2011social}.  We looked at the mobile phone log, counting the number of calls between each (directed) pair of individuals from October, 2010, to May, 2011.  The context of the study is a community of couples, around half of who have children, where one member of each couple is associated with a nearby major research university in North America.  Of the approximately 200 applicants, 130 actors of the network were selected in such a way as to represent the full community and sub-communities.  The entire community consists of 400 residents of a young family living community.  This study captured many aspects of the subjects beyond just the phone log, and among these we will look at religion and race.  For more details on the data and the collection process see \cite{aharony2011social}.  

The edges $y_{ijt}$ of the adjacency matrices $Y_t$ represents the number of phone calls from subject $i$ to subject $j$.  These counts were binned by month, and hence we had $T=8$ time points.  We eliminated any subjects who averaged less than one phone call, incoming or outgoing, per month.  This left 119 subjects.  Using counts rather than simply considering whether subject $i$ did or did not call subject $j$ during the $t^{th}$ month gives more insight into how gregarious each subject is, as well as better insight into how close each actor is to each other actor with whom they conversed via phone.

\begin{figure}[h]
\centering
\subfigure[$\beta_{IN}$]{
\includegraphics[width=7cm,height=4cm]{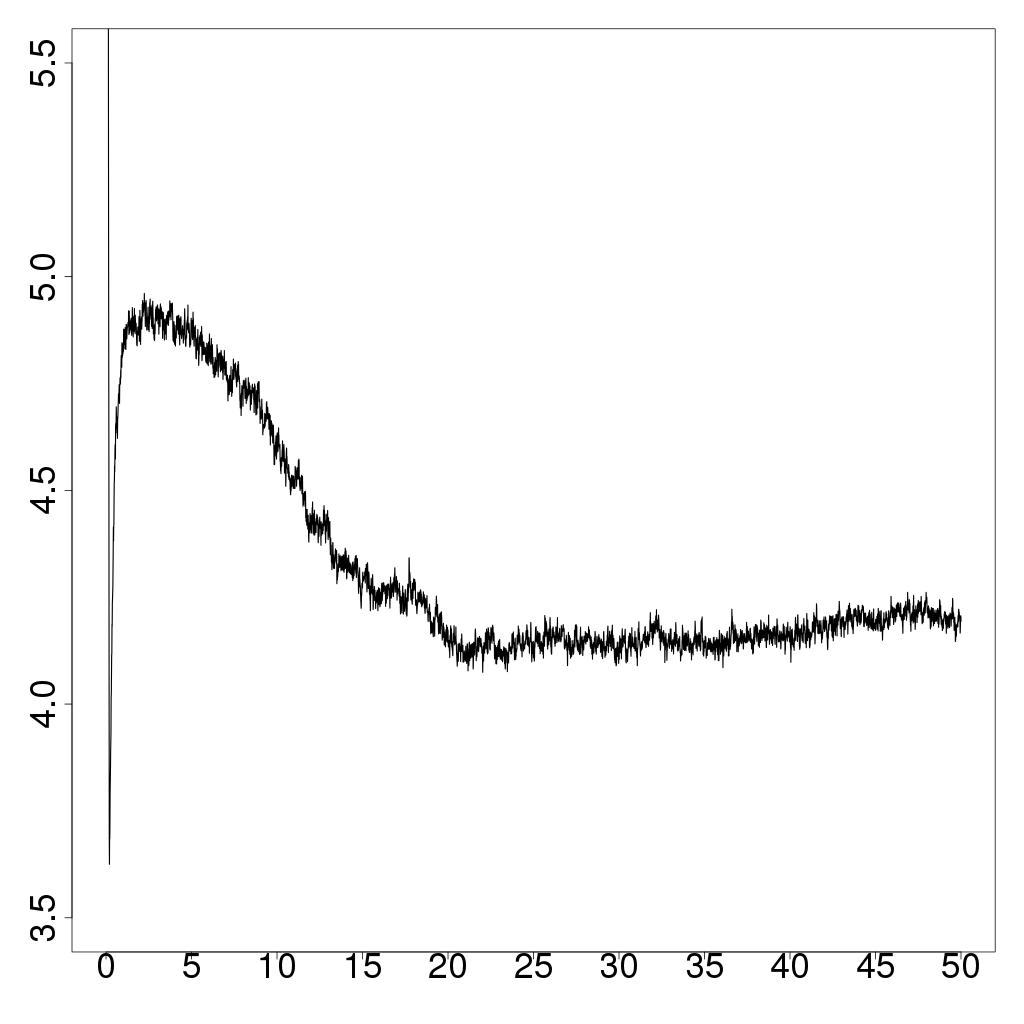}
}
\subfigure[$\beta_{OUT}$]{
\includegraphics[width=7cm,height=4cm]{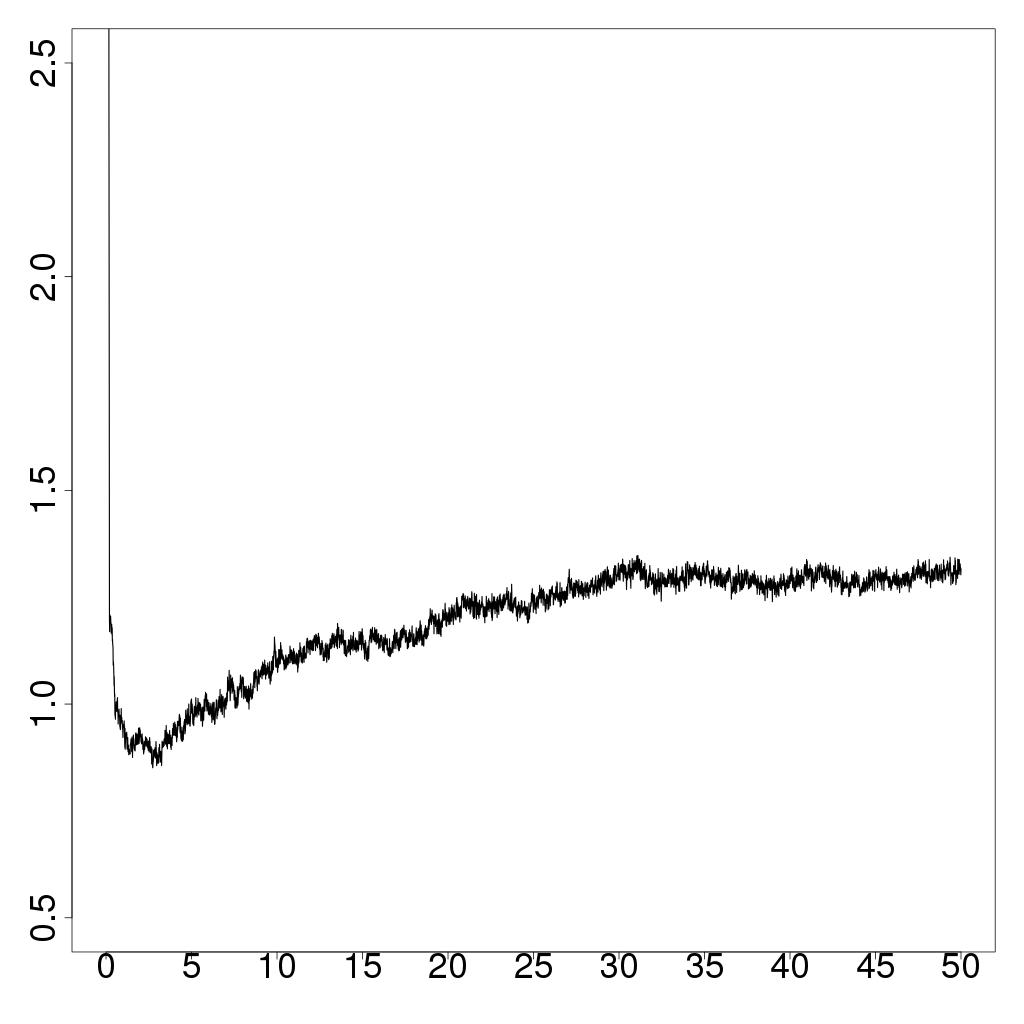}
}\\
\subfigure[$\tau^2$]
{
\includegraphics[width=7cm,height=4cm]{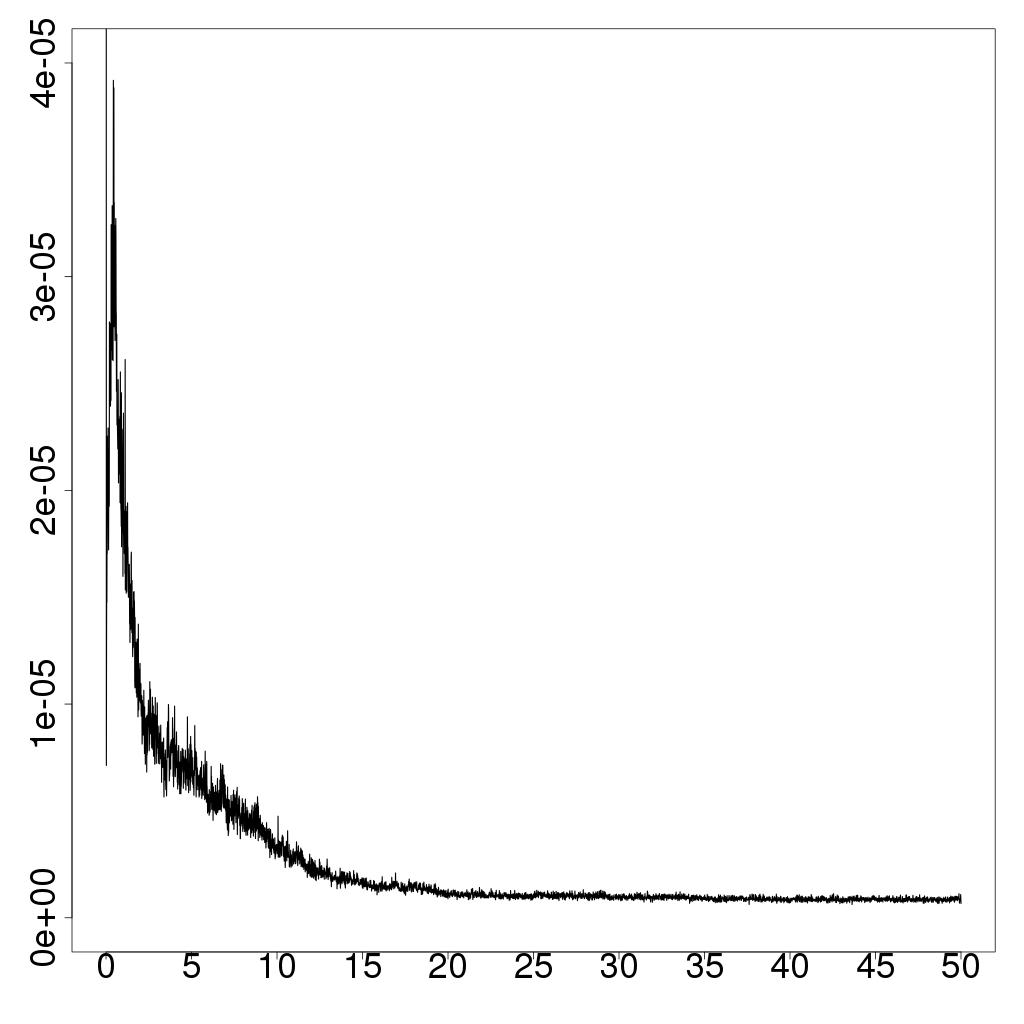}
}
\subfigure[$\sigma^2$]
{
\includegraphics[width=7cm,height=4cm]{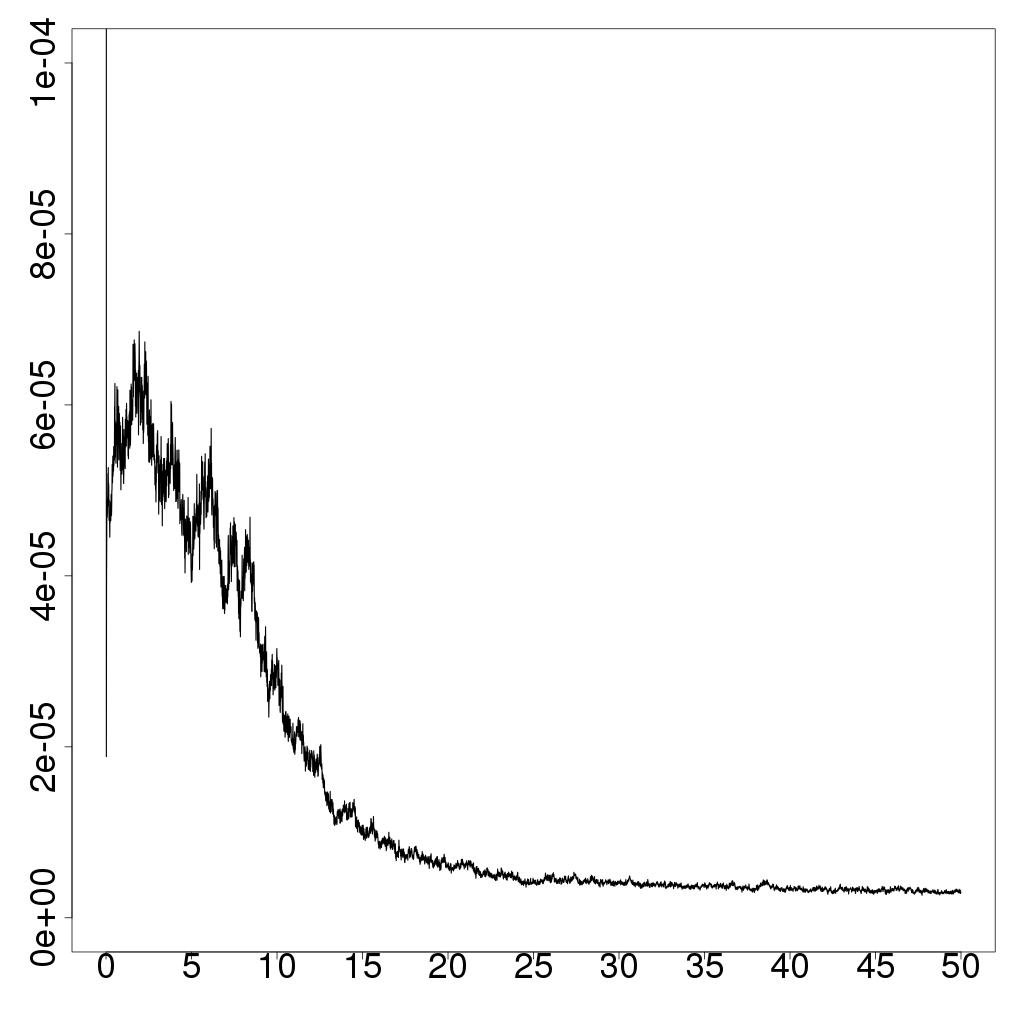}
}
\caption{MCMC trace plots for the model parameters corresponding to the Friends and Family data.  Horizontal axis is in iterations $\times10^4$.}
\label{FF_Trace}
\end{figure}

Initialization was performed according to Section \ref{Estimation}, setting $\delta=0.05$, $\sigma_0^2=1\times10^{-3}$, $\nu_{IN}=\nu_{OUT}=10$, and $\xi_{IN}=\xi_{OUT}=1000$.  We ran the MCMC algorithm until we obtained 500,000 samples, using a burnin of 300,000.  Figure \ref{FF_Trace} gives the trace plots  for $\beta_{IN}$, $\beta_{OUT}$, $\tau^2$ and $\sigma^2$; from this we can visually confirm that the MCMC chain has reached convergence.

The pseudo $R^2$ value was 0.819, implying a very good fit of the data. The posterior means of the coefficients were $\beta_{IN}=4.17$ and $\beta_{OUT}=1.29$, implying that, in the friendship network structure, popularity is more important than social activity.

\begin{figure}[p]
\centering
\subfigure[]
{
\includegraphics[width=5.5cm,height=5.5cm]{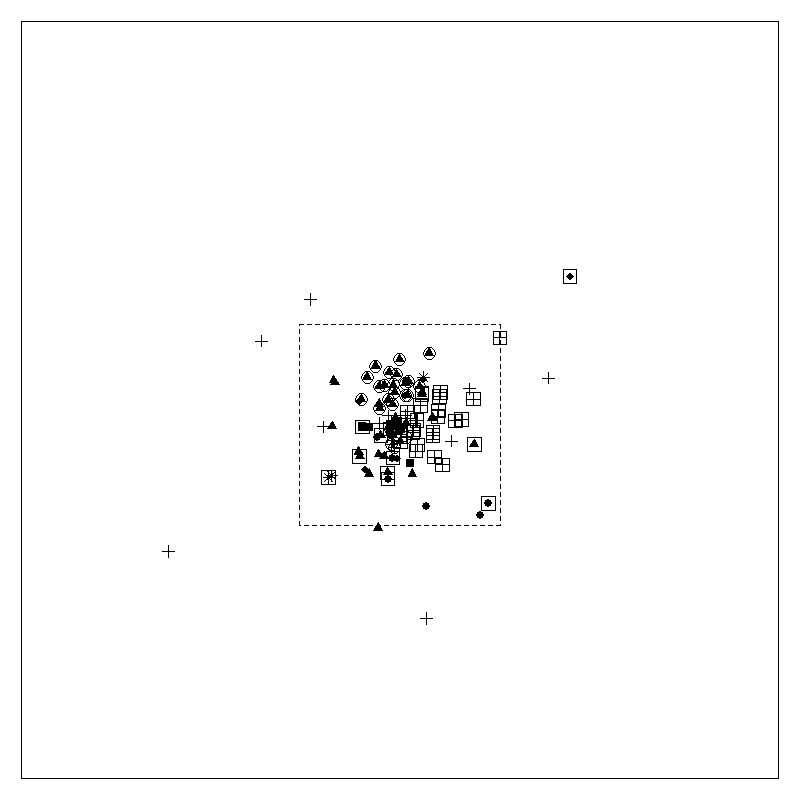}
}
\subfigure[]
{
\includegraphics[width=5.5cm,height=5.5cm]{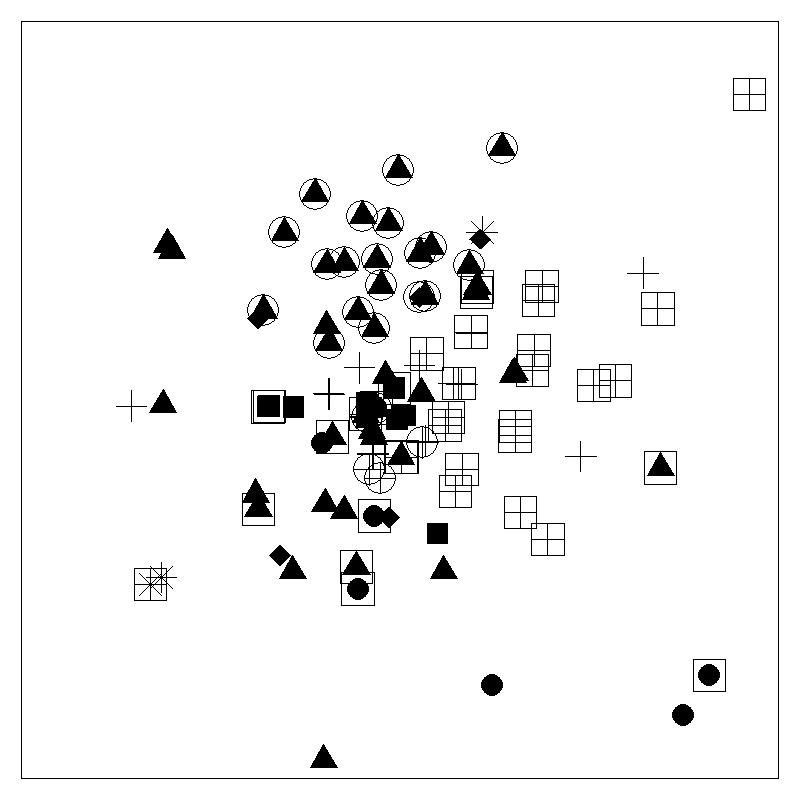}
}\\
\subfigure[]
{
\includegraphics[width=5.5cm,height=5.5cm]{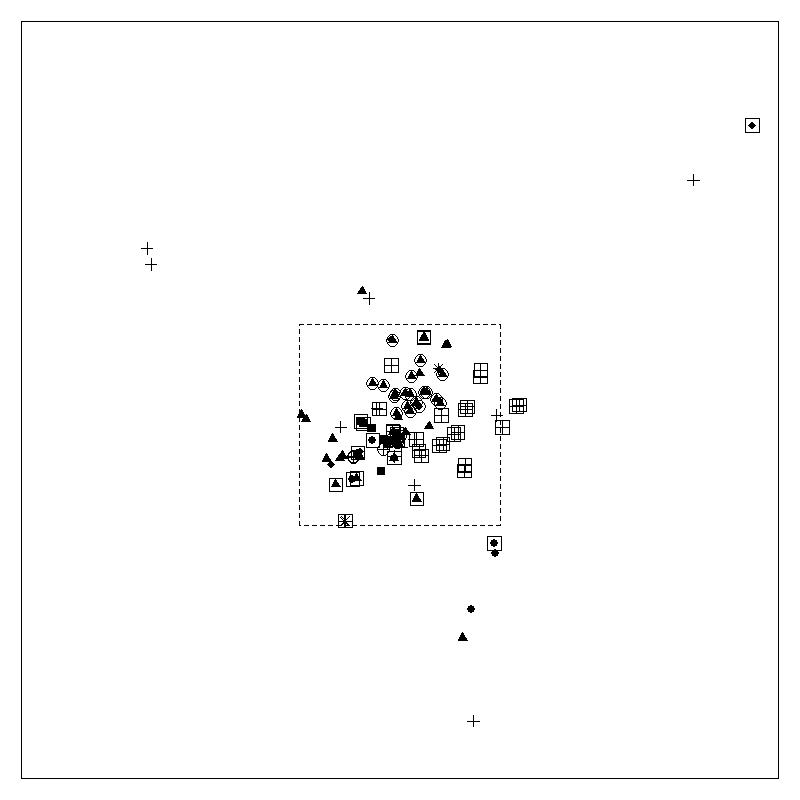}
}
\subfigure[]
{
\includegraphics[width=5.5cm,height=5.5cm]{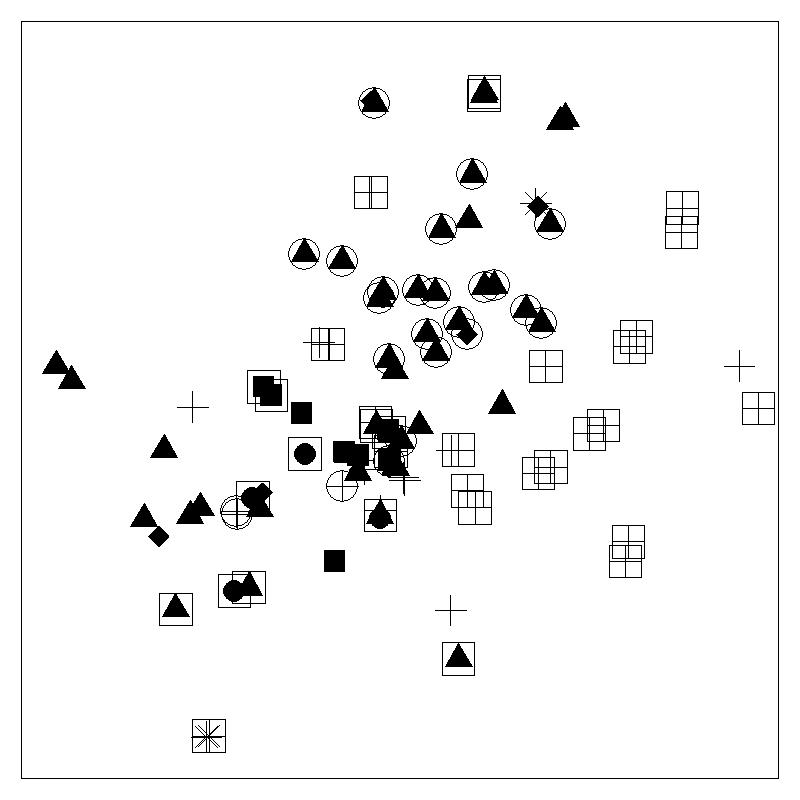}
}\\
\subfigure[]
{
\includegraphics[width=5.5cm,height=5.5cm]{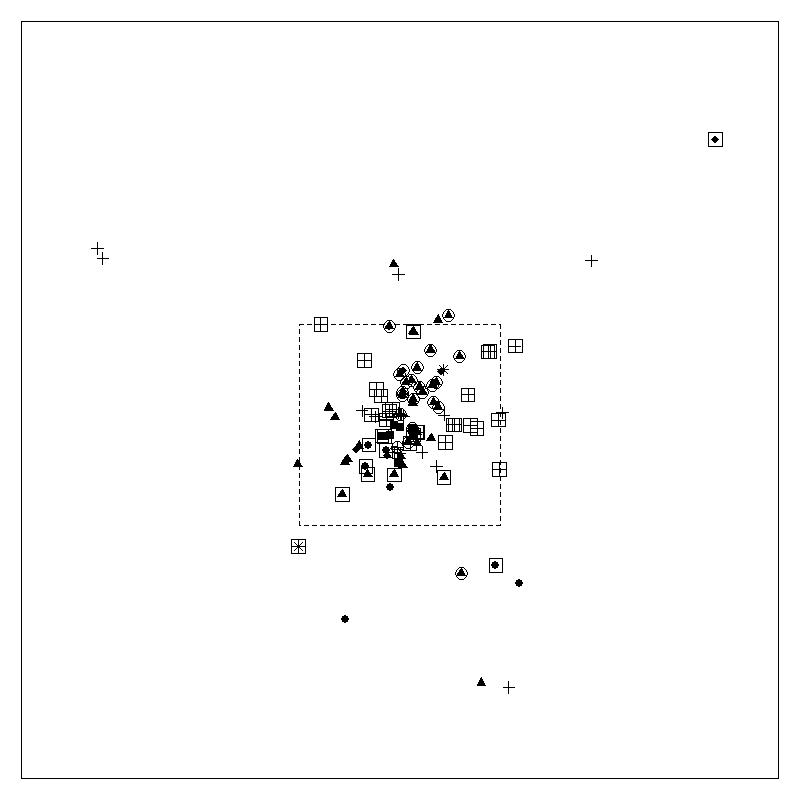}
}
\subfigure[]
{
\includegraphics[width=5.5cm,height=5.5cm]{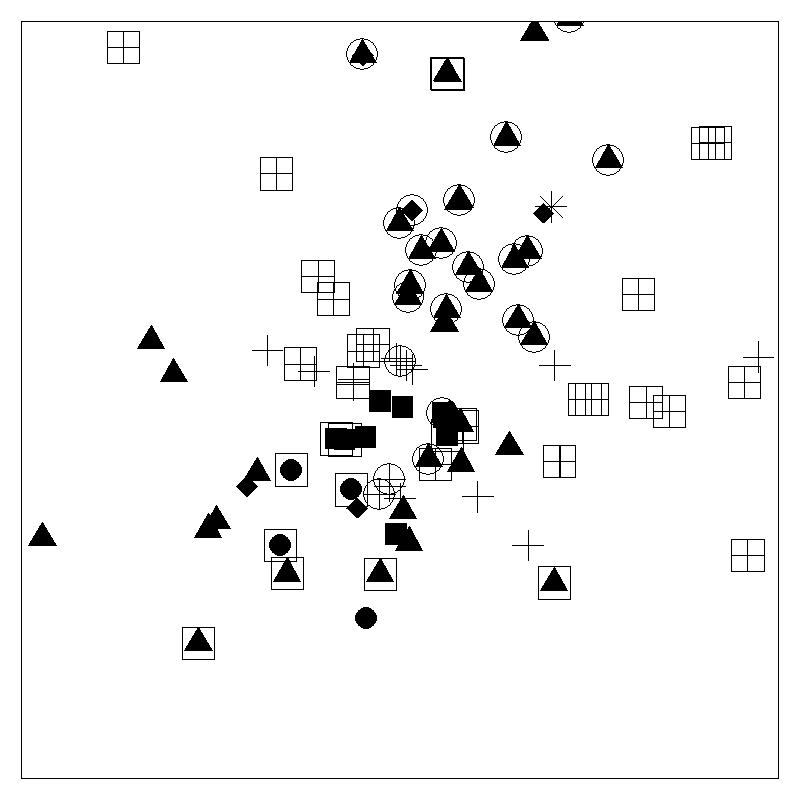}
}
\caption{Posterior means of latent positions for the Friends and Family data at times (a)-(b) 1, (c)-(d) 4, and (e)-(f) 8.  Each figure on the right is the zoomed in figure of the dotted box in the figures to their left.  Asians are indicated by +, blacks by asterisks, hispanics by solid squares, middle eastern by solid circles, and whites by solid triangles.  Actors that are boxed indicated that they are not at all religious, and actors that are circled indicated they are very religious.}
\label{FF_position}
\end{figure}

\begin{figure}
\centering
\subfigure[]{
\includegraphics[width=0.475\textwidth]{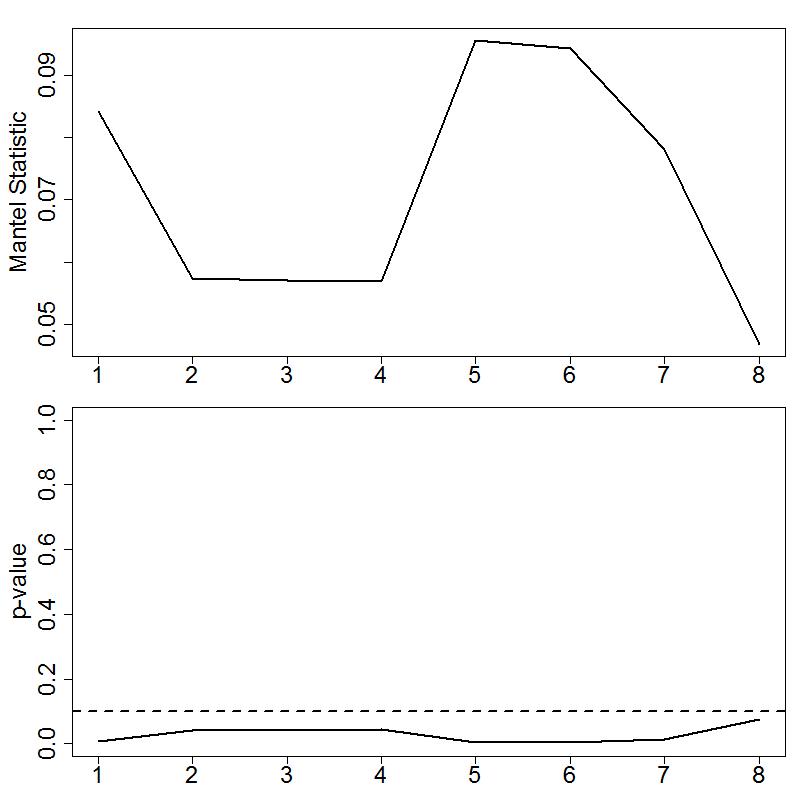}
}
\subfigure[]{
\includegraphics[width=0.475\textwidth]{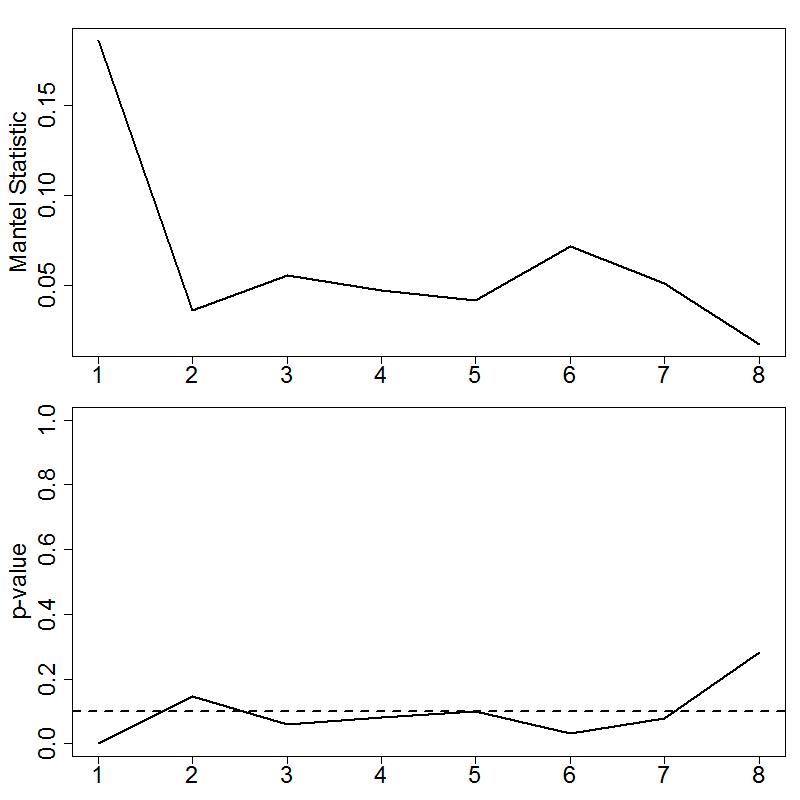}
}
\caption{Testing association between social position and (a) Race (b) Religion.  The top row is the test statistic, and the bottom row is the bootstrapped p-value.  The dotted line is 0.1.}
\label{Mantel}
\end{figure}

 Figure \ref{FF_position} gives a plot of the posterior mean latent positions at times 1, 3, 6, and 8.  The actors' shapes indicate the race (Asian, black, hispanic, middle eastern, or white), and the boxes or circles around the actors indicate their level of religion (either not at all religious or very religious).  From these figures we can see that there is some association between race and social position, as well as between religion and social position.  To verify this visual inspection, we performed a Mantel test between the posterior means of the latent positions and these two exogenous variables at each time point.  More specifically, we compared the distance matrix whose entries are given by $\|\widehat{\bx}_{it}-\widehat{\bx}_{jt}\|$, where $\widehat{\bx}_{it}$ is the posterior mean of $\bx_{it}$, to the distance matrix whose entries are 1 if actors $i$ and $j$ are not of the same race and 0 otherwise, as well as to the distance matrix whose entries are 1 if actors $i$ and $j$ are not of the same religious dedication.  The test statistic as well as the bootstrapped p-values are given in Figure \ref{Mantel}.  Thus from this analysis we have empirical evidence that one's social position is associated with race and religious dedication.

\subsection{World Trade Data}
\label{WorldTradeData}
World trade data, measuring annual exports/imports between countries in the years 1991-2000, was analyzed.  The data, given in millions of US dollars, was obtained through the Economic Web Institute at {\it http://www.economicswebinstitute.org/worldtrade.htm}, originally obtained through the IMF Direction of Trade (DOT) Yearbook.  Through this site, annual import/export data is available from 1948 to 2000.  We selected the most modern subset of this data which provided a reasonable number of countries that were present through all time points (e.g., not considering, e.g., states that become independent in the midst of the selected time period) as a pedagogical example.  The bilateral trade was measured in current millions of U.S. dollars; we analyzed the log of the trade amount, as is common in this context \citep[see, e.g., ][]{ward2013gravity}.  To account for global inflation/deflation and any other non-relational economic effects, the data was rescaled such that the total quantity of annual world trade is constant.  What we end up with then is a network consisting of 107 countries who were all involved in world trade through the 10 time years, 1991 to 2000, whose edges are non-negative reals.  For more information on the data see \cite{gleditsch2002expanded}.

\cite{ward2013gravity} developed a complex model for world trade data, combining a common economic model for world trade called the gravity model with aspects of the latent space model developed by \cite{hoff2005bilinear}.  Their approach uses one set of latent variables to model the incidence of trade and another set of latent variables to model the volume of trade.  However,  if we view the amount of trade between two countries as a positive-valued proxy indicating the strength of the relationship between the two countries' economies, our approach may be more appropriate.  Regardless, as the primary purpose of analyzing the world trade data described above is to serve as a pedagogical example of our more general methodology, we have maintained the more simple model framework of \cite{SewellChen14} with our extension for weighted network data, demonstrating the data augmentation scheme of Section \ref{NonNegativeContinuousEdges}.

\begin{figure}[htb]
\centering
\subfigure[$\beta_{IN}$]
{
\includegraphics[width=5cm,height=3cm]{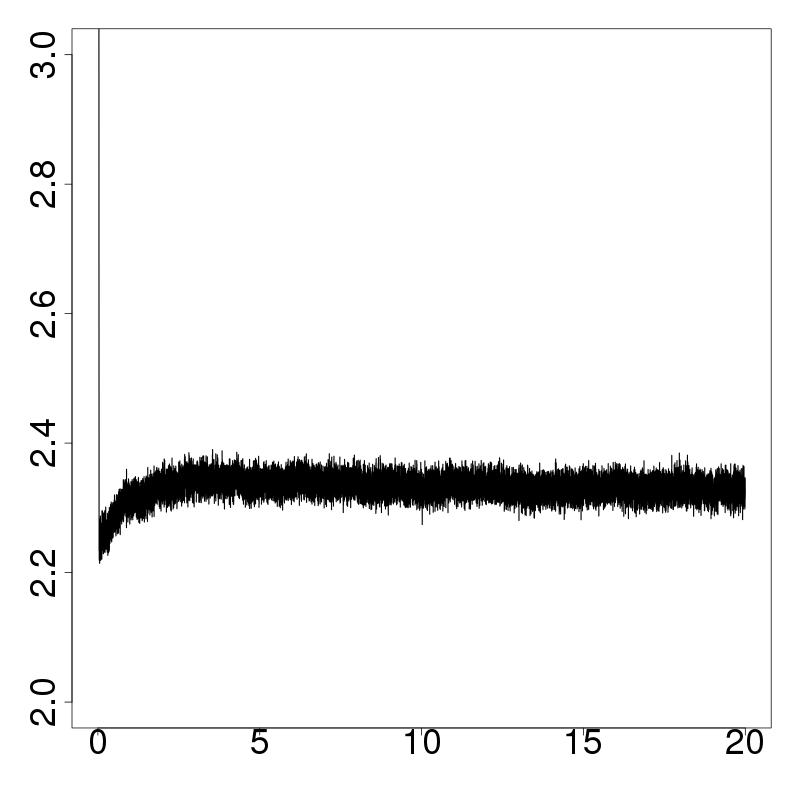}
}
\subfigure[$\beta_{OUT}$]
{
\includegraphics[width=5cm,height=3cm]{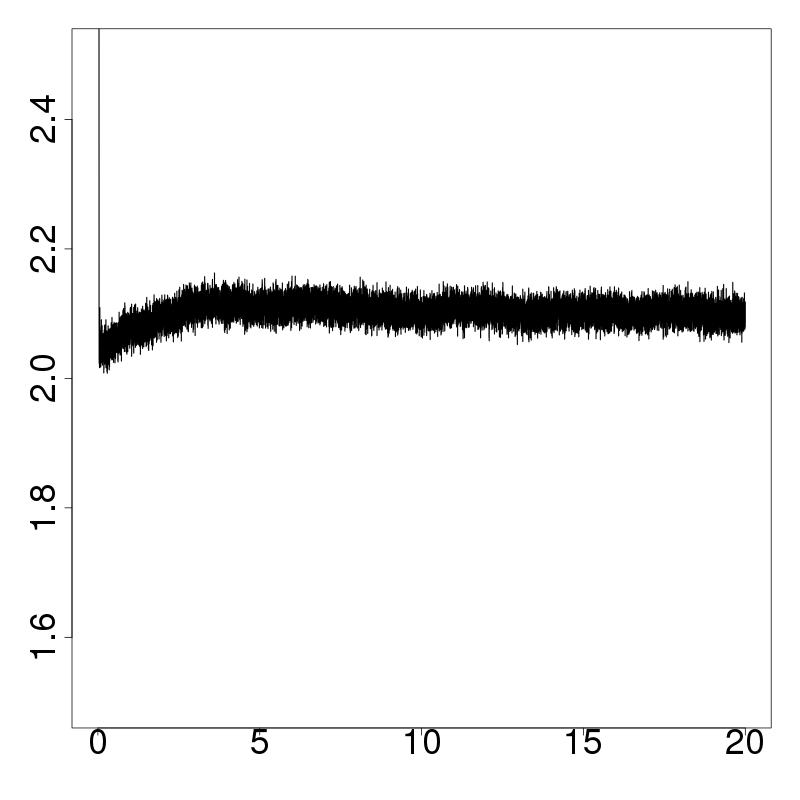}
}
\subfigure[$\sigma^2$]
{
\includegraphics[width=5cm,height=3cm]{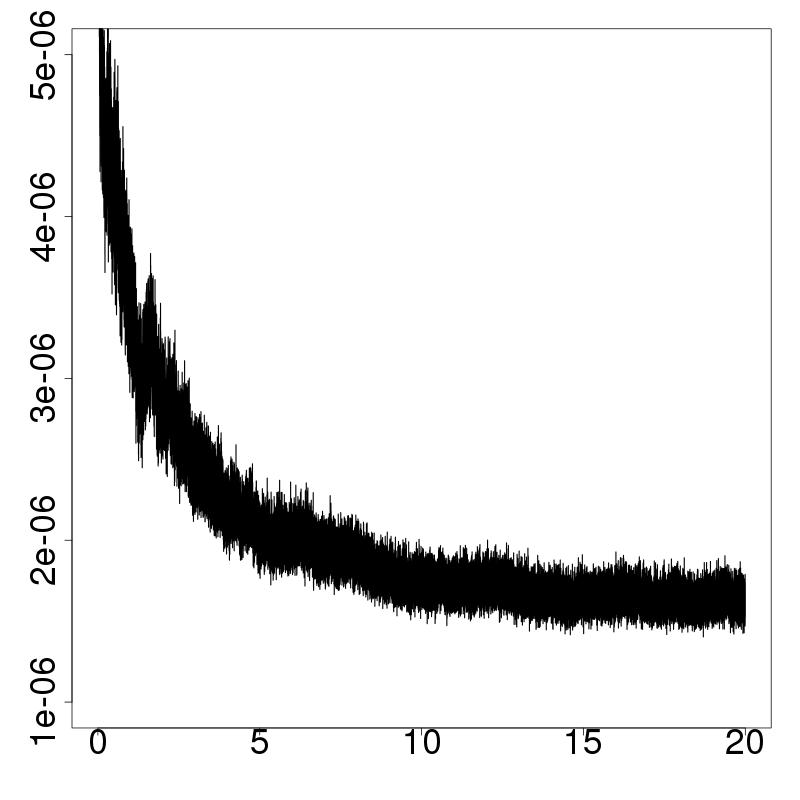}
}
\subfigure[$\tau^2$]
{
\includegraphics[width=5cm,height=3cm]{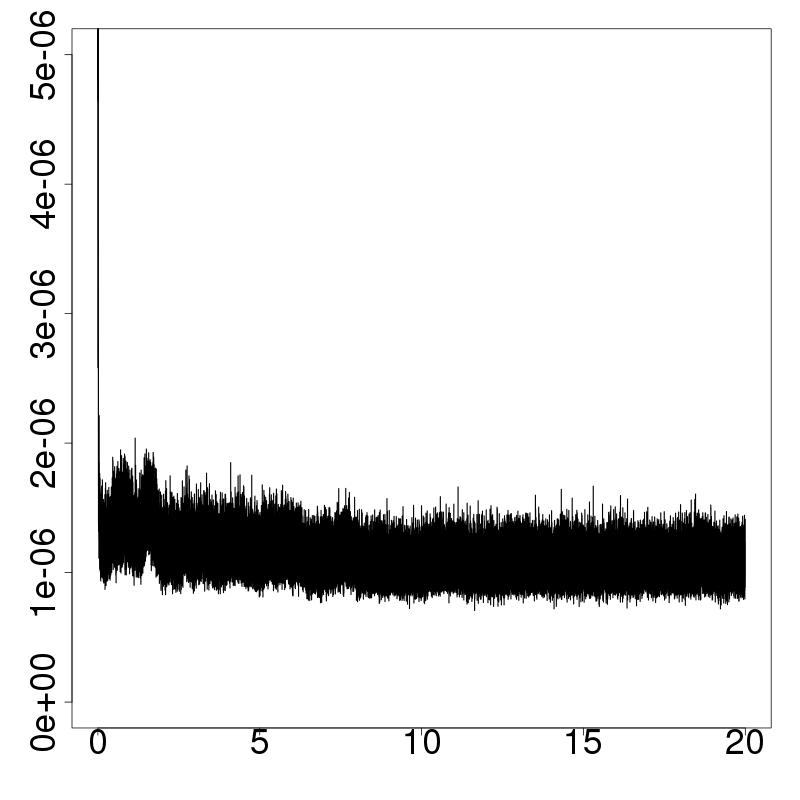}
}
\subfigure[$\gamma^2$]
{
\includegraphics[width=5cm,height=3cm]{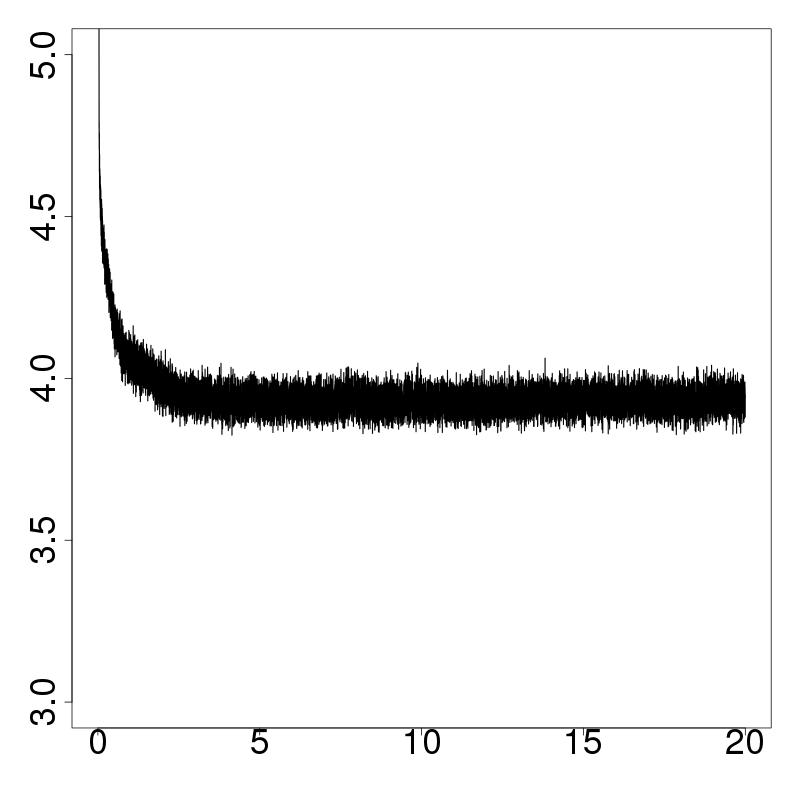}
}
\caption{MCMC trace plots for the model parameters corresponding to the world trade data.  Horizontal axis is in iterations $\times 10^4$.}
\label{worldTrade_trace}
\end{figure}

The hyperparameters for the priors of $\sigma^2$, $\tau^2$, $\gamma^2$, and $\br$ were formulated according to the description in Section 3.  We set $\delta=$, $\sigma^2_0=1\times 10^{-3}$, $\gamma_0^2=25$, $\beta_{IN}=\beta_{OUT}=10$, and $\nu_{IN}=\nu_{OUT}=1000$.  Figure \ref{worldTrade_trace} gives the trace plots for $\beta_{IN}$, $\beta_{OUT}$, $\sigma^2$, $\tau^2$, and $\gamma^2$.  A burn-in of 125,000 was used, leaving a chain of length 75,000; from this we can visually confirm that the MCMC chain has reached convergence.

The pseudo $R^2$ value was 0.970, indicating a very good fit of the data.  The estimates for $\beta_{IN}$ and $\beta_{OUT}$ were 2.33 and 2.10 respectively, implying that the amount of trade is determined more by the importing country than the exporting country, but only slightly so.  If $\beta_{IN}$ had been much larger than $\beta_{OUT}$ then this would have suggested that the importer was in larger control of the trade relationship, and if $\beta_{OUT}$ was much larger than $\beta_{IN}$ we would say the same about the exporter.  However, in our case we see that the two coefficients are close to each other, suggesting that the trade relationship is closely balanced.

\begin{figure}[p]
\centering
\includegraphics[height=7cm,width=15cm]{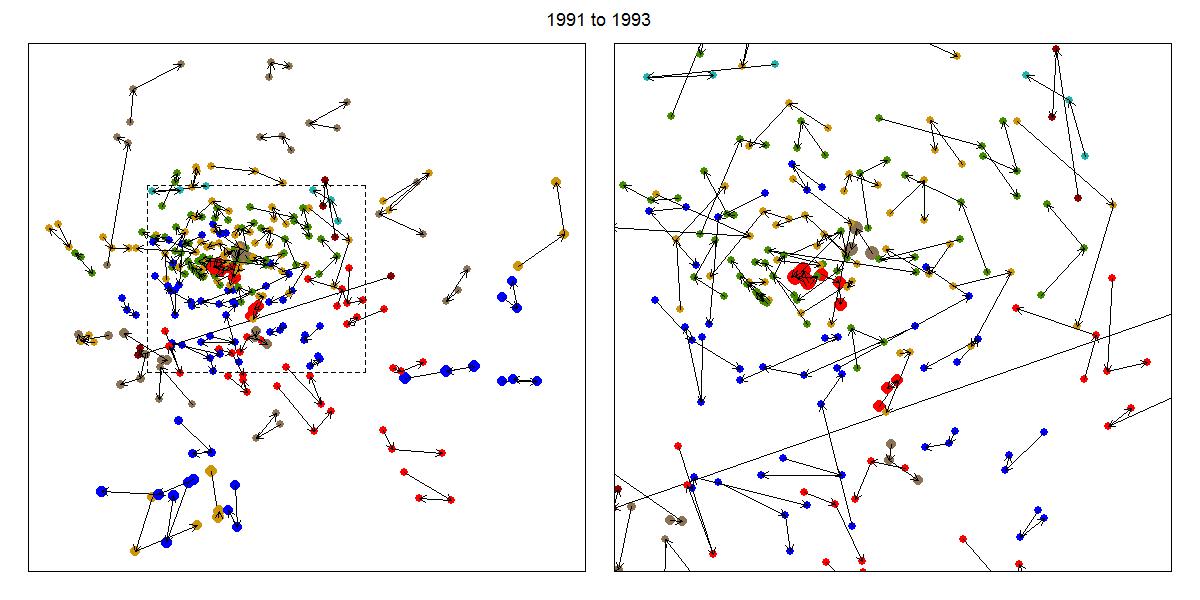}
\includegraphics[height=7cm,width=15cm]{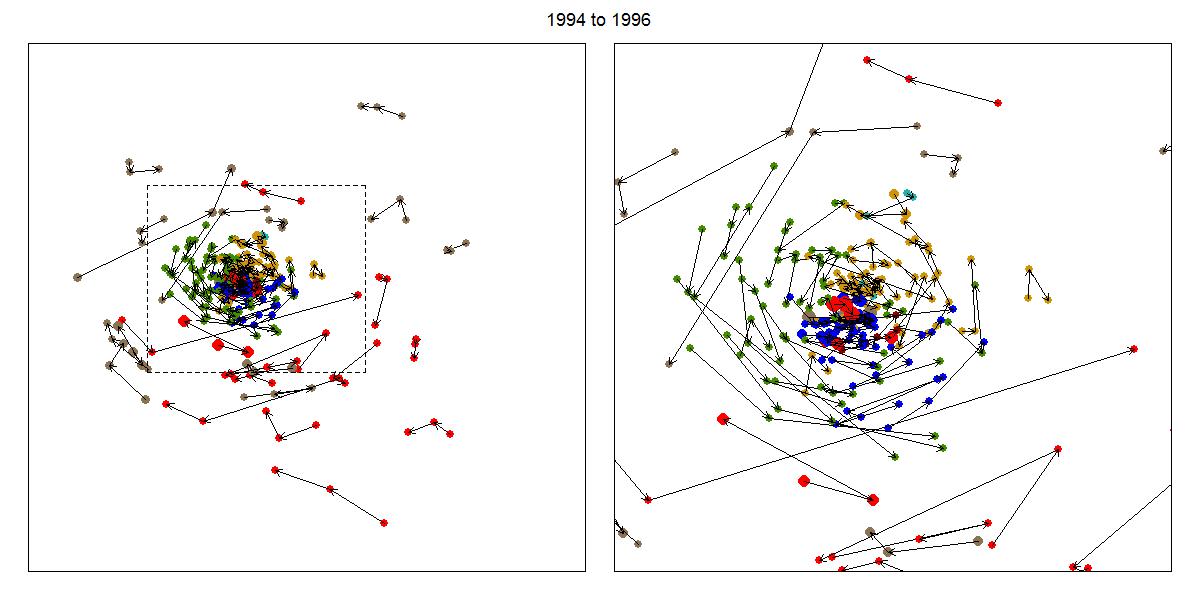}
\includegraphics[height=7cm,width=15cm]{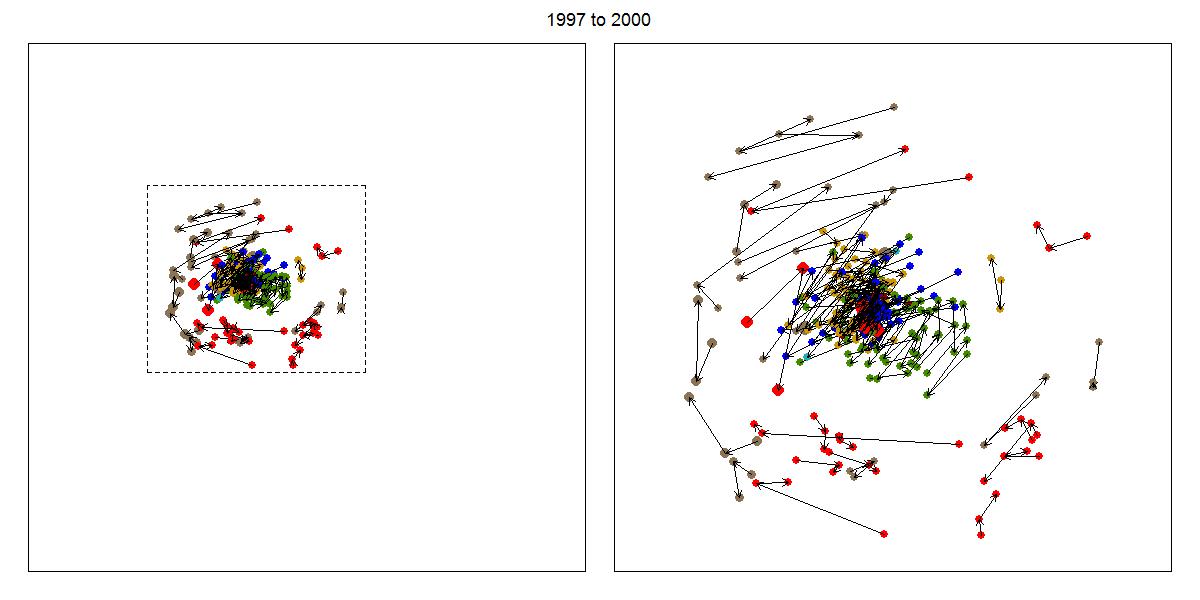}
\caption{Latent positions of each nation from the world trade import/export relational data.  Each figure on the right is the zoomed in figure of the dotted box in the figures to their left.  See main text for key to colors.}
\label{worldTrade_positions}
\end{figure}

Figure \ref{worldTrade_positions} gives plots of the posterior mean latent positions, broken up into three time periods: from 1991 to 1993, from 1994 to 1996, and from 1997 to 2000.  Temporal direction is shown via arrows.  The size of the actor corresponds to its $r_i$ value.  Each color represents a geographical region, where green is Africa, yellow is Asia, dark red is Eurasia, blue is Europe, red is North America and the Caribbean, sea green is Oceania, and brown is South America.  It is apparent that the actors move within the latent space much less during each of these three periods than during the transition from 1993 to 1994 and from 1996 to 1997.  These two major shrinkage events occurring both coincide with major events in world trade.  In 1993, the General Agreement on Tariffs and Trade was updated, which would later lead to the creation of the World Trade Organization (WTO) (see http://www.wto.org).  Looking at Figure \ref{worldTrade_positions}, we can see that there is already some shrinkage happening during the year 1993 which then continues going into 1994.  Specifically we see that certain continents (Africa, Asia, and Europe) come together during this time.  Europe is arguably the clearest case, and it turns out there is a good reason for this: the European Economic Area was established on January 1, 1994.  Regarding the second shrinkage event in 1997, a publication from the WTO states that ``the volume of world merchandise exports grew by 9.5 per cent in 1997.”  This is seen visually in Figure \ref{worldTrade_positions}.  However, since the original data had been scaled to account for such growth, we conclude that the reason for this growth in world exports is not due to existing relationships getting stronger, but rather to the formation of many more trading relationships.

\section{Discussion}
\label{Discussion}
Using the weights associated with edges makes better use of data than only incorporating the existence or non-existence of an edge.  The weighted data is more informative and should lead to more accurate inference.  It also eliminates the need to make an arbitrary user-defined cutoff for determining whether an edge should be one or zero.

We have described a general strategy for applying the latent space model for dynamic networks to data consisting of weighted edges.  This can be applied either directly or indirectly through additional latent variables.  We have demonstrated the flexibility of the latent space models for dynamic networks by modeling cosponsorship count data and non-negative continuous world trade data.

Our latent space models can handle directed edges of many edge types, model both local and global structures, inherently account for transitivity, and yield a rich visualization of the data.  An additional note is that using the MH within Gibbs sampling allows edges missing at random or missing completely at random to be incorporated into the model and estimated \citep[see][for details]{SewellChen14}.

\section*{Appendix: Full Conditional Distributions}
The full conditional distributions for $\tau^2$ and $\sigma^2$ are respectively
\begin{equation}
\pi(\tau^2|{\cal X}_1) \sim  \mbox{IG}(2+\delta+np/2,(1+\delta)\tau_0^2+\frac{1}{2}\sum_{i=1}^n\|{\bf X}_{i1}\|^2), \label{tau2conddist}
\end{equation}
\begin{equation}
\pi(\sigma^2|{\cal X}_1,{\cal X}_2,\ldots,{\cal X}_T) \sim \mbox{IG}(2+\delta+\frac{np(T-1)}{2},(1+\delta)\sigma_0^2+\frac{1}{2}\sum_{t=2}^T\sum_{i=1}^n\|{\bf X}_{it}-{\bf X}_{i(t-1)}\|^2), \label{sigma2conddist}
\end{equation}
for $\delta$, $\tau_0^2$, $\sigma_0^2 >0$.

We let $\pi_{y_{ijt}}\triangleq\pi(y_{ijt}|{\cal X}_t,\boldsymbol\Psi)$ as given in (\ref{count_Lik1}) and (\ref{count_Lik2}) if we have count edges, or as given in (\ref{tobit_Lik}) if we have non-negative real edges.  Then the full conditional distribution for ${\bf X}_{it}$ in these two cases is
\begin{equation}
\pi({\bf X}_{it}|Y_{1:T},\boldsymbol\Psi)
\propto \left\{\begin{array}{ll}
\left( \prod\limits_{j\neq i} \pi_{y_{ijt}}\pi_{y_{jit}} \right)\cdot N({\bf X}_{it}|{\bf 0},\tau^2 I_p)\cdot N({\bf X}_{i(t+1)}|{\bf X}_{it},\sigma^2 I_p), & \mbox{ if $t=1$} \\
\left( \prod\limits_{j\neq i} \pi_{y_{ijt}}\pi_{y_{jit}} \right)\cdot N({\bf X}_{i(t+1)}|{\bf X}_{it},\sigma^2 I_p)\cdot N({\bf X}_{it}|{\bf X}_{i(t-1)},\sigma^2 I_p), & \mbox{ if $1<t<T$}\\
\left( \prod\limits_{j\neq i} \pi_{y_{ijt}}\pi_{y_{jit}} \right)\cdot N({\bf X}_{it}|{\bf X}_{i(t-1)},\sigma^2 I_p),& \mbox{ if $t=T$}.
\end{array}\right.
\label{Xktconddist}
\end{equation}

The full conditional distribution for each of the model parameters follows the form
\begin{equation}
\pi(\psi|Y_{1:T},{\cal X}_{1:T},\boldsymbol\Psi\backslash\{\psi\}) \propto \left[ \prod_{t=1}^T \pi(Y_t|{\cal X}_t,\boldsymbol\Psi) \right]\cdot \pi(\psi)
\end{equation}
where $\boldsymbol\Psi\backslash\{\psi\}$ is the set of parameters excluding $\psi$, and for count data $\psi \in\{\beta_{OUT}, \beta_{IN}, \boldsymbol{r}\}$ and for non-negative real data $\psi \in\{\beta_{OUT}, \beta_{IN}, \gamma^2, \boldsymbol{r}\}$.

\section*{Acknowledgements}
We thank the referees for their valuable ideas and suggestions which have led to the improvement of this paper.  This work was supported by {\bf Fill in this part}.

\bibliographystyle{asa}
\bibliography{extension1}

\begin{thebibliography}{31}
\newcommand{\enquote}[1]{``#1''}
\expandafter\ifx\csname natexlab\endcsname\relax\def\natexlab#1{#1}\fi

\bibitem[{Aharony et~al.(2011)Aharony, Pan, Ip, Khayal, and
  Pentland}]{aharony2011social}
Aharony, N., Pan, W., Ip, C., Khayal, I., and Pentland, A. (2011),
  \enquote{Social fMRI: Investigating and shaping social mechanisms in the real
  world,} \textit{Pervasive and Mobile Computing}, 7, 643--659.

\bibitem[{Barrat et~al.(2005)Barrat, Barth{\'e}lemy, and
  Vespignani}]{barrat2005effects}
Barrat, A., Barth{\'e}lemy, M., and Vespignani, A. (2005), \enquote{The effects
  of spatial constraints on the evolution of weighted complex networks,}
  \textit{Journal of Statistical Mechanics: Theory and Experiment}, 2005,
  P05003.

\bibitem[{Cameron and Windmeijer(1996)}]{cameron1996r}
Cameron, A.~C. and Windmeijer, F.~A. (1996), \enquote{R-squared measures for
  count data regression models with applications to health-care utilization,}
  \textit{Journal of Business \& Economic Statistics}, 14, 209--220.

\bibitem[{Durante and Dunson(2014)}]{durante2014nonparametric}
Durante, D. and Dunson, D.~B. (2014), \enquote{Nonparametric Bayes dynamic
  modelling of relational data,} \textit{Biometrika}, 101, 125--138.

\bibitem[{Gleditsch(2002)}]{gleditsch2002expanded}
Gleditsch, K.~S. (2002), \enquote{Expanded trade and GDP data,} \textit{Journal
  of Conflict Resolution}, 46, 712--724.

\bibitem[{Globerson et~al.(2004)Globerson, Chechik, Pereira, and
  Tishby}]{globerson2004euclidean}
Globerson, A., Chechik, G., Pereira, F., and Tishby, N. (2004),
  \enquote{Euclidean embedding of co-occurrence Data,} \textit{Advances in
  Neural Information Processing Systems}, 17, 497--504.

\bibitem[{Hanneke et~al.(2010)Hanneke, Fu, and Xing}]{hanneke2010discrete}
Hanneke, S., Fu, W., and Xing, E.~P. (2010), \enquote{Discrete temporal models
  of social networks,} \textit{Electronic Journal of Statistics}, 4, 585--605.

\bibitem[{Hoff(2005)}]{hoff2005bilinear}
Hoff, P.~D. (2005), \enquote{Bilinear mixed-effects models for dyadic data,}
  \textit{Journal of the American Statistical Association}, 100, 286--295.

\bibitem[{Hoff(2011)}]{hoff2011hierarchical}
--- (2011), \enquote{Hierarchical multilinear models for multiway data,}
  \textit{Computational Statistics \& Data Analysis}, 55, 530--543.

\bibitem[{Krause et~al.(2003)Krause, Frank, Mason, Ulanowicz, and
  Taylor}]{krause2003compartments}
Krause, A.~E., Frank, K.~A., Mason, D.~M., Ulanowicz, R.~E., and Taylor, W.~W.
  (2003), \enquote{Compartments revealed in food-web structure,}
  \textit{Nature}, 426, 282--285.

\bibitem[{Krivitsky(2012)}]{krivitsky2012exponential}
Krivitsky, P.~N. (2012), \enquote{Exponential-family random graph models for
  valued networks,} \textit{Electronic Journal of Statistics}, 6, 1100--1128.

\bibitem[{Krivitsky and Butts(2012)}]{krivitsky2012rank}
Krivitsky, P.~N. and Butts, C.~T. (2012), \enquote{Exponential-family random
  graph models for rank-order relational data,} \textit{arXiv:1210.0493}.

\bibitem[{Krivitsky and Handcock(2014)}]{krivitsky2014separable}
Krivitsky, P.~N. and Handcock, M.~S. (2014), \enquote{A separable model for
  dynamic networks,} \textit{Journal of the Royal Statistical Society: Series
  B}, 76, 29--46.

\bibitem[{Kunegis et~al.(2009)Kunegis, Lommatzsch, and
  Bauckhage}]{kunegis2009slashdot}
Kunegis, J., Lommatzsch, A., and Bauckhage, C. (2009), \enquote{The slashdot
  zoo: mining a social network with negative edges,} in \textit{Proceedings of
  the 18th International Conference on World Wide Web}, ACM, pp. 741--750.

\bibitem[{McKelvey and Zavoina(1975)}]{mckelvey1975statistical}
McKelvey, R.~D. and Zavoina, W. (1975), \enquote{A statistical model for the
  analysis of ordinal level dependent variables,} \textit{Journal of
  Mathematical Sociology}, 4, 103--120.

\bibitem[{Morgan(2014)}]{morgan2014latent}
Morgan, J. (2014), \enquote{The Latent path model for dynamic networks,}
  \textit{Ohio State University Manuscript}.

\bibitem[{Newman(2004)}]{newman2004analysis}
Newman, M.~E. (2004), \enquote{Analysis of weighted networks,} \textit{Physical
  Review E}, 70, 056131.

\bibitem[{Olgu{\i}n et~al.(2009)Olgu{\i}n, Gloor, and
  Pentland}]{olguin2009capturing}
Olgu{\i}n, D.~O., Gloor, P.~A., and Pentland, A.~S. (2009), \enquote{Capturing
  individual and group behavior with wearable sensors,} in \textit{Proceedings
  of the 2009 AAAI Spring Symposium on Human Behavior Modeling, SSS}, vol.~9.

\bibitem[{Onnela et~al.(2007)Onnela, Saram{\"a}ki, Hyv{\"o}nen, Szab{\'o},
  De~Menezes, Kaski, Barab{\'a}si, and Kert{\'e}sz}]{onnela2007analysis}
Onnela, J.-P., Saram{\"a}ki, J., Hyv{\"o}nen, J., Szab{\'o}, G., De~Menezes,
  M.~A., Kaski, K., Barab{\'a}si, A.-L., and Kert{\'e}sz, J. (2007),
  \enquote{Analysis of a large-scale weighted network of one-to-one human
  communication,} \textit{New Journal of Physics}, 9, 179.

\bibitem[{Opsahl et~al.(2010)Opsahl, Agneessens, and Skvoretz}]{opsahl2010node}
Opsahl, T., Agneessens, F., and Skvoretz, J. (2010), \enquote{Node centrality
  in weighted networks: Generalizing degree and shortest paths,} \textit{Social
  Networks}, 32, 245--251.

\bibitem[{Opsahl and Panzarasa(2009)}]{opsahl2009clustering}
Opsahl, T. and Panzarasa, P. (2009), \enquote{Clustering in weighted networks,}
  \textit{Social Networks}, 31, 155--163.

\bibitem[{Raftery et~al.(2012)Raftery, Niu, Hoff, and Yeung}]{raftery2012fast}
Raftery, A.~E., Niu, X., Hoff, P.~D., and Yeung, K.~Y. (2012), \enquote{Fast
  inference for the latent space network model using a case-control approximate
  likelihood,} \textit{Journal of Computational and Graphical Statistics}, 21,
  901--919.

\bibitem[{Sarkar and Moore(2005)}]{sarkar2005dynamic}
Sarkar, P. and Moore, A. (2005), \enquote{Dynamic social network analysis using
  latent space models,} \textit{ACM SIGKDD Explorations Newsletter}, 7, 31--40.

\bibitem[{Sarkar et~al.(2007)Sarkar, Siddiqi, and Gordon}]{sarkar2007latent}
Sarkar, P., Siddiqi, S.~M., and Gordon, G.~J. (2007), \enquote{A latent space
  approach to dynamic embedding of co-occurrence data,} in \textit{Proceedings
  of the Eleventh International Conference on Artificial Intelligence and
  Statistics AISTATS}, vol.~7.

\bibitem[{Sewell and Chen(2015{\natexlab{a}})}]{SewellChen14rank}
Sewell, D.~K. and Chen, Y. (2015{\natexlab{a}}), \enquote{Analysis of the
  formation of the structure of social networks using latent space models for
  ranked dynamic networks,} \textit{Journal of the Royal Statistical Society:
  Series C}, DOI: 10.1111/rssc.12093.

\bibitem[{Sewell and Chen(2015{\natexlab{b}})}]{SewellChen14}
--- (2015{\natexlab{b}}), \enquote{Latent space models for dynamic networks,}
  \textit{Journal of the American Statistical Association}, DOI:
  10.1080/01621459.2014.988214.

\bibitem[{Veall and Zimmermann(1994)}]{veall1994practitioners}
Veall, M.~R. and Zimmermann, K.~F. (1994), \enquote{Goodness of fit measures in
  the Tobit model,} \textit{Oxford Bulletin of Economics and Statistics}, 56,
  485--499.

\bibitem[{Ward et~al.(2013)Ward, Ahlquist, and Rozenas}]{ward2013gravity}
Ward, M.~D., Ahlquist, J.~S., and Rozenas, A. (2013), \enquote{Gravity's
  rainbow: A dynamic latent space model for the World Trade Network,}
  \textit{Network Science}, 1, 95--118.

\bibitem[{Xing et~al.(2010)Xing, Fu, and Song}]{xing2010state}
Xing, E.~P., Fu, W., and Song, L. (2010), \enquote{A state-space mixed
  membership blockmodel for dynamic network tomography,} \textit{The Annals of
  Applied Statistics}, 4, 535--566.

\bibitem[{Yang and Knoke(2001)}]{yang2001optimal}
Yang, S. and Knoke, D. (2001), \enquote{Optimal connections: strength and
  distance in valued graphs,} \textit{Social Networks}, 23, 285--295.

\bibitem[{Zhang and Horvath(2005)}]{zhang2005general}
Zhang, B. and Horvath, S. (2005), \enquote{A general framework for weighted
  gene co-expression network analysis,} \textit{Statistical Applications in
  Genetics and Molecular Biology}, 4, Article 17.

\end{thebibliography}

\end{document}